\documentclass[aps,amsmath,twocolumn,amssymb,floatfix,showpacs,
superscriptaddress,footinbib,longbibliography,prb,reprint]{revtex4-1}
\pdfoutput=1
\usepackage{graphics}
\usepackage{bm}
\usepackage{epsfig}
\usepackage{enumerate}
\usepackage{subfigure}
\usepackage{amsmath}
\usepackage{color}
\usepackage{braket}
\usepackage{graphicx}
\usepackage{hyperref}
\hypersetup{
    colorlinks=true,
    linktoc=page,
    linkcolor=red,
    citecolor=blue,
    urlcolor=magenta
}
\usepackage{soul}
\sethlcolor{red}

\begin{document}

\title{Topological energy pumping in a quasi-periodically driven four-level system}

\author{Vansh Kaushik}
\email{vkaushik@sissa.it}
\affiliation{SISSA, via Bonomea 265, 34136 Trieste, Italy}

\author{Sayan Choudhury}
\email{ sayanchoudhury@hri.res.in}
\affiliation{Harish-Chandra Research Institute, a CI of Homi Bhabha National Institute, Chhatnag Road, Jhunsi, Allahabad 211019}

\author{Tanay Nag}
\email{ tanay.nag@hyderabad.bits-pilani.ac.in, Corresponding author}
\affiliation{Department of Physics, BITS Pilani-Hyderabad Campus, Telangana 500078, India}

\begin{abstract}
We investigate a quasi-periodically driven four-level system that serves as a temporal analog of topological phenomena found in four-band models with intertwined spin and orbital degrees of freedom. Under a two-tone drive in the strong-driving regime, the system realizes a two-dimensional synthetic Floquet lattice, thus facilitating the realization of topological energy pumping.  For a temporal quantum spin Hall insulator, we find that the rates of emission and absorption of energy between the two drives are not exactly opposite for a given band. However, when contributions from two chiral symmetric partner bands are added, they become exactly opposite. This quantized rate of energy exchange is a direct consequence of propagating edge modes in the real-space model, which we further characterize by computing the spin-Chern number. Interestingly, our analysis yields zero rate of exchange of energy between the drives for a temporal higher-order topological insulator, suggesting the presence of localized corner modes that we characterize by the mid-gap Wannier spectra.  {Our findings uncover the role of chiral, particle-hole and time reversal symmetries on the energy dynamics in temporal quantum spin Hall and higher-order topological insulators.} Finally, we demonstrate that the perfect (imperfect) nature of the fidelity during the time-evolution of the system serves as a characteristic signature of a topological (trivial) phase.

\end{abstract}

\maketitle

\section{Introduction}

The discovery of the integer quantum Hall effect, and its theoretical linkage between quantized Hall conductance and the topological invariant of Thouless, Kohmoto, Nightingale, and den Nijs, revolutionized quantum condensed matter physics by introducing the concept of topology \cite{KlitzingPRL1980,TKNNPRL1982}. 
It has been found that the topological phases of matter are symmetry-protected states and hence they lie outside the Landau paradigm. Depending upon the symmetry and dimensionality of the system, several celebrated topological models have been extenesively analyzed; these include 
topological insulators \cite{hasan2010colloquium,qi2011topological}, Chern insulators \cite{HaldanePRL1988}, quantum spin-Hall insulators (QSHI) \cite{kane2005quantum,kane2005quantumPRL,Saha21}, and the Bernevig, Hughes and Zhang (BHZ) model \cite{BHZPRL2006,BernevigBOOK}. The topological phases in these systems are characterized by an appropriate invariant. For instance, the Chern (spin-Chern) number is used to distinguish between topological and trivial phases in the Chern (QSH) insulator  \cite{TKNNPRL1982,FuKane2007,fu2007topological,MoorePRB2007,BernevigBOOK}. Interestingly, the bulk topological invariants are connected with the presence of boundary modes~\cite{FuKane2007,RahulRoyPRB2009,MoorePRB2007,fu2007topological,FuTimereversal2006,TurnerCrystallinePRB2012,fu2011topological}; this is known as the bulk boundary correspondence. In particular, the presence of
charge and spin-polarized edge modes, lying inside the bulk gap, in Chern insulators and QSHI are captured by the Chern number and spin-Chern number respectively~\cite{Saha21,nag2025extended,Sheng06}. The bulk boundary correspondence has been generalized in the context of higher-order topological insulator (HOTI) where the $d$-dimensional topological system supports $(d-n)$-th dimensional boundary modes in a $n$-th order topological phase with $n>1$ and are characterized by mid-gap Wannier spectra \cite{benalcazar2017,benalcazarprb2017,Song2017,Langbehn2017,schindler2018,Franca2018,Ezawaphosphorene,wang2018higher,Ezawakagome,Geier2018,Khalaf2018,ezawa2019second,luo2019higher,Roy2019,RoyGHOTI2019,Trifunovic2019,Szumniak2020,Ni2020,BiyeXie2021,trifunovic2021higher}. Note that two-dimensional ($d=2$) topological model such as QSHI, one can obtain both first order topological insulator (FOTI), and HOTI phases with one-dimensional edge modes and zero-dimensional corner modes, respectively,  under appropriate FOT and HOT mass terms. We note that these  HOTI phases have been experimentally realized  \cite{schindler2018higher,Experiment3DHOTI.VanDerWaals,Aggarwal2021}.

Concurrently with developments in topological physics, rapid advances have been made in exploring the non-equilibrium dynamics of many-body systems. In particular, external periodic drives have emerged as a powerful tool for engineering and controlling many-body systems~\cite{oka2019floquet,Eckardt17, Meinert16,wang2013observation,rechtsman2013photonic,gangopadhay2025counterdiabatic}. Following the framework of Floquet theory~\cite{Shirley65,bukov2015universal}, it has been shown that periodic drive induces topology without having any prior static analog~\cite{Oka09,Kitagawa10,Gu11,lindner2011floquet,Lindner13,Goldman14,Rudner13,Kundu14,Lababidi14,Mondal23,Mondal23b}. Interestingly, Floquet enegineering techniques have already been employed for realizing the HOT phase \cite{Nag19c,Huang2020,Zhu21,ghosh2023generation,Arnob22}.
The introduction of interaction and disorder adds further richness to these systems, enabling the realization of Floquet many-body localization~\cite{abanin2016theory,Zhang16} and Floquet time crystal phases \cite{khemani2016phase,Else16,yao2017discrete,else2020discrete,zaletel2023colloquium,huang2018clean,Lyu2020eternal}. 
Finally, we note that the catalog of Floquet phases can be considerably extended by employing multi-tone drive protocols where various orders such as charge order, topological order, and electric polarization are engineered ~\cite{Wang23,Nag19b,ikeda2022floquet,Viebahn21,Mizuta23,banerjee2025exact,dutta2025controlling}.

In recent years, the dynamics of aperiodically driven systems have received increasing attention~\cite{VerdenyPuigMintert16,PhysRevX.7.031034,Nag21,Long22,kundu2024statistical,PhysRevLett.126.040601,Zhao22,Guanghui23,Tiwari25,anisur2025quasi,dutta2025prethermalization,ghosh2025heating}. These investigations have led to the discovery of novel non-equilibrium phases of matter such as Floquet time spirals~\cite{Hongzheng19} and time quasicrystals~\cite{dumitrescu2018logarithmically,he2025experimental} in quasi-periodically driven systems,  and time rondeau crystals in many-body systems subjected to random multipolar driving~\cite{zhao2023temporal,kumar2024stable,moon2024experimental,ma2025stable}. Intriguingly, the dynamics of a $d-$dimensional quantum system driven by a continuous quasi-periodic drive composed of $D$- incommensurate tones can be mapped to a $d+D$-dimensional model in a composite synthetic space ~\cite{Crowley19,Zihao21,Peng18,Psaroudaki23,Nathan22,Lantagne24,Zihao24,long2022coupled,Long22,korber2020interacting}. Consequently, a two-tone driven two-level system (qubit) can be employed to realize the half-BHZ model representing some variant of quantum anomalous Hall insulator~\cite{Martin17,boyers2020exploring,crowley2020half,sridhar2024quantized,vuina2023giant}; in this case, the topological (trivial) phase is characterized by quantized (non-quantized) energy pumping between the drives. Importantly, in this context, Majorana multiplexing, time reflection and refraction, frequency conversion, and quasi-Floquet prethermalization are also studied extensively \cite{Guanghui23,Martin17,Lantagne24,Zihao21,Zihao24,Crowley19,sridhar2024quantized,Psaroudaki23,Long23,Park23,Nathan22,schmid2025self}. 
Given the above background and employing the Boltzmann transport formalism \cite{ashcroft1976solid,giuliani2008quantum}, we proceed to investigate the dynamics of a quasi-periodically driven system that can realize the temporal analog of the  QSHI and HOTI phases. In particular, we seek answers for the following unaddressed questions: (a) What is the nature of energy pumping between the drives for the temporal QSHI model?, (b) How is the energy dynamics modified in the presence of a HOT mass term?, and (c) What is the connection between the bulk topological invariant and energy pumping?

In this work, we address these aforementioned questions by investigating the energy dynamics of a four-level system (a ququart) that is driven by two incommensurate frequencies. This system can realize the temporal analog of a four-band model, where the interplay between spin and orbital degrees of freedom plays a crucial role. Our analysis reveals that the rates of energy absorption and emission are not exactly symmetric for a given band. This is markedly different compared to the quasi-periodically driven single-spin model where energy absorption and emission rates are  exactly opposite for both the bands. Interestingly however, for QSHI phase the energy absorption and emission rates become completely opposite,  when we add the contributions from two different bands, related by chiral symmetry and particle-hole symmetry. The regular quantized (irregular non-quantized) rate of energy-pumping is directly related to the quantized (non-quantized) spin-Chern number characterizing the underlying propagating topological edge modes (trivial bulk localized mode) of the real-space model. Next, we extend this analysis to the temporal HOTI model where corner localized modes of the real-space model are manifested in the zero absorption and emission of energy after adding both the chiral partner bands. This is further established by the mid-gap Wannier spectra characterizing the HOTI nature of the phase.  { Our results reveal how chiral, particle-hole, and time-reversal symmetries govern the energy dynamics in temporal quantum spin Hall and higher-order topological insulators.} Finally, we demonstrate that the perfect (fluctuating) nature of the time-evolved fidelity serves as a distinct signature of a topological (trivial) phase.  

The paper is structured as follows: In Sec. \ref{s2}, we introduce the model Hamiltonian for both temporal QSHI and HOTI, discuss the Floquet formalism, and the corresponding observables that we study.  In Sec. \ref{s3}, we analyze the observables namely, the energy exchange rates,  spin-Chern number, Wannier spectra, and providing a comprehensive understanding of the topological transitions associated with FOTI and HOTI phases. We examine the time evolution of fidelity for both the models and connect this behavior to the topological or trivial nature of the system.  Finally, we conclude with a summary of our findings and possible future directions in Sec. \ref{s4}.

\section{Model and observables}
\label{s2}

\subsection*{Model}
\label{s2ss1}

To investigate the energy pumping dynamics for a two-tone driven ququart, we adapt the two-dimensional QSHI model where the momentum modes can be considered as the Fourier modes in the frequency space. Motivated by the QSHI, the Hamiltonian in the presence of a quasi-periodic drive is described by \cite{Nag19,BHZPRL2006,bernevig2006quantum}
\begin{equation}
    H^{\rm IC}_{\text{QSHI}} = \sum_{i=1}^{2} T_i S_i -  T_3 \left( m - \sum_{i=1}^{2} C_i \right),
    \label{eq:ham_qshi}
\end{equation}
where 
and \(S_i = \sin(\omega_i t + \phi_i)\), \(C_i = \cos(\omega_i t + \phi_i)\) with incommensurate drive frequencies $\omega_{1,2}$, initial phases $\phi_{1,2}$, and $t$ denotes time. We consider $\omega_2=\beta \omega_1$ with $\beta=(\sqrt{5}+1)/2$.
The four-component mutually anti-commuting Hermitian matrices are given by $T_1 = \sigma_3  \tau_1$, $T_2 = \sigma_0 \tau_2$, and $T_3 = \sigma_0 \tau_3$, where $\bm {\sigma}$ and $\bm{\tau}$ represent the standard Pauli matrices  {denoting  spin and orbital} degrees of freedom, respectively, without loss of generality.  This model is directly connected to the QSHI when 
$\omega_i t + \phi_i$ is replaced   {by $k_i b_i$} with $i=x,y$, and the   {lattice spacing $b_i$ is taken to be unity, $b_i=1$}.  {In the context of a tight-binding model, one can anticipate the $S_i$ terms as spin-orbit coupling while $C_i$ terms as hopping.} We note that in general, the incommensurate drive protocol maps the temporal QSHI model to a real-space model with irregular lattice spacing. Note that QSHI phases are found in the amorphous materials as well as quasi-crystals without spatial periodicity
\cite{Huang18,Fritz18}. We emphasize however, that in this work, we perform our analyses in the strong driving regime, where the irregularity of the lattice does not play a significant role as the concept of a  band structure remains valid
~\cite{Martin17}.

For completeness, we now discuss the underlying  {tight-binding QSHI  model where the  Bloch momenta ${\bm k}=(k_x,k_y)$ are used to examine the band topology. } 
The QSHI model exhibits two distinct first-order-topological (FOT) phases with spin-Chern numbers $(C_{\uparrow},C_{\downarrow})=(-1,1)$ and $(1,-1)$ for $-2<m<0$ and $0<m<2$, respectively while $C_{\alpha}$ represents the Chern number for $\alpha$ spin degree of freedom. These phases are characterized by four zero-energy helical edge states under open boundary conditions, where  
up- and down-spins can execute clockwise [counter-clockwise] and counter-clockwise [clockwise] motion along the edges for $C_{\uparrow},C_{\downarrow}=(-1,1)$ [$(1,-1)$], respectively. For $|m|>2$, the QSHI model hosts a trivial phase where states are localized inside the bulk.  { The QSHI phase is protected by time-reversal symmetry $T=i \sigma_2  \tau_0 K$ ($K$ denotes the complex conjugation), the chiral symmetry $C=\sigma_2  \tau_1$ and particle-hole symmetry $P=\sigma_3  \tau_1 K$ are also respected. As a result, the, QSHI Hamiltonian follows the following constraints $TH({\bm k})T^{-1}=H(-{\bm k})$, $CH({\bm k})C^{-1}=-H({\bm k})$, and $PH({\bm k})P^{-1}=-H(-{\bm k})$, and $T^2=-I$.}

Recently, it has been shown that QSHI model can be elevated to a HOTI when the Wilson-Dirac mass term is added. In particular, a HOTI described by $H_{\rm HOTI}=Q_{\rm QSHI}+ \Delta (C_1 - C_2) T_4 $ with $T_4 = \sigma_1  \tau_1$ hosts four zero-energy corner modes while the edge states are gapped out \cite{Nag19,Roy2019}. This phenomenon originates from the Jackiw-Rebbi theorem that allows for the corner modes to be present at the intersections of the edges, while the edge states are gapped out due to the term $(C_1 - C_2)$ that yields a sign-changing profile   {across vanishing Wilson-Dirac mass term with $k_x=k_y$  which represents the diagonal direction in $xy$ plane}. Interestingly, while the Wilson-Dirac mass term breaks the time reversal and four-fold rotational symmetry, the corner modes are protected by chiral and particle-hole symmetries.  {$H_{\rm HOTI}$ follows the following two constraints only $CH({\bm k})C^{-1}=-H({\bm k})$, and $PH({\bm k})P^{-1}=-H(-{\bm k})$.} The corner modes are characterized by the mid-site Wannier center, which in turn is manifested by mid-gap ($0.5$) eigenvalues of the Wannier spectrum.

Motivated by the above observation, we extend the temporal QSHI model to a temporal HOTI model:
\begin{equation}
    H^{\rm IC}_{\text{HOTI}} = H^{\rm IC}_{\text{QSHI}} + \Delta \big(\cos (\omega_1 t + \phi_1) - \cos (\omega_2 t + \phi_2) \big) T_4. 
    \label{eq:ham_hoti}
\end{equation}
where the $k_i$ replaces $\omega_i t + \phi_i$ in the HOTI model. The models discussed above (Eqs. (\ref{eq:ham_qshi}) and  (\ref{eq:ham_hoti})) exhibit qualitatively different kinds of energy dynamics that we will explore below.   {Note that, similar to the tight-binding model, the temporal  Wilson-Dirac mass term changes its sign across the line $\omega_1 t + \phi_1=\omega_2 t + \phi_2$. }   
Likewise, in analogy with amorphous QSHIs, amorphous HOTI phases also exist \cite{Agarwala20,Yang_2024} in the absence of a periodic lattice structure. However, as stated earlier, we work in the strong drive limit allowing the band picture to survive \cite{Martin17}.

 {We now discuss the 
symmetry aspects of the temporal QSHI and HOTI models in the presence of incommensurate drive. Without loss of generality, the off-set phase $\phi_{1,2}$ can be considered to be zero for the symmetry analysis. 
Given the correspondence  ${\bm k} \to {\bm \omega} t$, between tight-binding and temporal model, we find that $P H^{\rm IC}_{\rm X}(t) P^{-1}= -H^{\rm IC}_{\rm X}(-t)$, $C H^{\rm IC}_{\rm X}(t) C^{-1}= -H^{\rm IC}_{\rm X}(t)$, and $T H^{\rm IC}_{\rm QSHI}(t) T^{-1}= H^{\rm IC}_{\rm QSHI}(-t)$ with X=QSHI, HOTI. Importantly, $T H^{\rm IC}_{\rm HOTI}(t) T^{-1} \ne H^{\rm IC}_{\rm HOTI}(-t)$.}

\subsection*{Floquet prescription}
\label{s2ss2}

In the case of two commensurate frequencies, one can obtain the infinite-dimensional Floquet Hamiltonian in the frequency space by integrating time over a full period $T$. This procedure breaks down however, when the two frequencies become incommensurate. Nevertheless, we can still employ the following anstaz for the time-evolution of any initial state : $|\psi(t)\rangle = \sum_{n} \exp(-i\bm{n} \cdot \bm{\omega} t)|\psi_{n}(t)\rangle$~\cite{Martin17}, where 
the time-evolved wavefunction $|\psi(t)\rangle$ is expanded in terms of Fourier harmonics $|\psi_{n}(t)\rangle$ such that 
 $e^{\pm i \omega_i t}= | n_i \mp 1\rangle \langle n_i|$.
Here $n_i$ represents the photon number of the $i-$th drive~\cite{long2021nonadiabatic}, and in the present context, this directly translates into the index of Floquet lattice. Adapting this form into the 
time-dependent Schr\"odinger equation $ i \frac{d|\psi\rangle}{dt} = H |\psi\rangle$, we obtain the following Fourier space $H_{\rm HOTI}^{\rm IC}$ Floquet lattice model
\begin{equation}
\begin{aligned}
    i \partial_t |\psi_{n_1,n_2}\rangle &= \frac{1}{2} \left( i  T_1 -  T_3  + \Delta T_4\right) e^{i \phi_1} \ket{\psi_{\bm{n_{-},n}}} \\
    &\quad + \frac{1}{2} \left( -i  T_1 -  T_3  + \Delta T_4 \right) e^{-i \phi_1} \ket{\psi_{\bm {n_{+},n}}}  \\
    &\quad + \frac{1}{2} \left( i  T_2 -  T_3 - \Delta T_4\right) e^{i \phi_2} \ket{\psi_{\bm {n,n'_-}}} \\
    &\quad + \frac{1}{2} \left( -i  T_2 -  T_3 - \Delta T_4\right) e^{-i \phi_2} \ket{\psi_{\bm {n,n'_+}}} \\
    &\quad + \left( m T_3   - n_1 \omega_1    - n_2 \omega_2 \right ) \ket{\psi_{\bm {n,n}}}.
\end{aligned}
\label{eq:ham_fqshi}
\end{equation}
with ${\bm n}= (n_1,n_2)$, ${\bm n_{\pm}}=(n_1 \pm1,n_2)$, and ${\bm n'_{\pm}}=(n_1 ,n_2 \pm1)$.  
In the above expression, $(n_1,n_2)$ can be thought of $x$ and $y$ indices for a frequency space lattice, and the physical interpretation of  ${n}=(n_1,n_2)$ denotes the number of photons in the drives.  Importantly,  the first four terms containing $\ket{\psi_{n_{\pm},n}}$ and $\ket{\psi_{n, n'_{\pm}}}$ represent hopping along $x$ and $y$ direction in the frequency lattice, respectively. The last term having $\ket{\psi_{n,n}}$ 
denotes the on-site potential term. The  first part with $m T_3$ in this on-site term represents a uniform on-site potential in the Floquet lattice, while the second and third parts denote   {on-site potential}.that correspond to `electric fields' along $x$ and $y$  directions in the Floquet lattice, respectively.

The aforementioned Floquet lattice Hamiltonian can be decomposed in the form $H_{\rm HOTI}^{\rm IC}=H_{\bm{q}} \hat{n}_{\bm q}+ H_{\bm{n}} \hat{n}_{\bm n} $~\cite{Zihao24,Peng18,Martin17,Crowley19} where $H_{\bm{q}}$ represents the 
Fourier-transformed 'momentum-space' form of the Hamiltonian, in the absence of an electric field i.e., force-free, containing only the hopping and mass terms:
\begin{equation}
\begin{aligned}
    H_{\bm{q}}(\Delta) &=  T_1 \sin(q_1 + \phi_1) + T_2 \sin(q_2 + \phi_2) \\
    &\quad -  T_3 \big[m - \cos(q_1 + \phi_1) - \cos(q_2 + \phi_2)\big] \\
    & \quad + \Delta T_4\big [ \cos(q_1 + \phi_1) - \cos(q_2 + \phi_2) \big] \\
    &= {\bm H_{q_1}} \cdot {\bm T} + {\bm H_{q_2}} \cdot {\bm T} + {\bm H_{0}} \cdot {\bm T},
\end{aligned}
\label{eq:ham_pesudoq}
\end{equation}
where ${\bm H_{q_1}}= (\sin (q_1 + \phi_1), 0, \cos(q_1 + \phi_1),\cos(q_1 + \phi_1))$, 
${\bm H_{q_2}}= (0,\sin (q_2 + \phi_2), \cos(q_2 + \phi_2),-\cos(q_2 + \phi_2))$, ${\bm H}_0=(0,0,-m,0)$, ${\bm T}=(T_1, T_2, T_3, T_4)$, and $\hat{n}_{\bm q}$ is the occupation of the momentum state ${\bm q} = \{q_1,,q_2\}$. 
Furthermore, $H_{\bm n}$ represents the `real space' Hamiltonian representing the effect of a force:  
\begin{equation}
    H_{\bm{n}} = -\bm{n} \cdot \bm{\omega},
\label{eq:ham_force}    
\end{equation}
and $\hat{n}_{\bm n}$ is the occupation of site ${\bm n} = \{n_1,n_2\}$. Interestingly, the time-evolution of the momentum is governed by this force: ${\bm q}= {\bm \omega} t$. Finally, we note that our analysis also applies to the QSHI Hamiltonian, $H_{\rm QSHI}^{\rm IC}$ which is the $\Delta=0$ limit of the HOTI model. 

We now proceed to work in the strong driving regime, where the band gaps of $H_{\bf q}$ is much larger than the driving frequencies, $\omega_1, \omega_2$. In this limit, we can assume that the pseudo-momentum forms bands of energy $\epsilon_{\bm q}$, and the force term in the Floquet lattice, ${\bm E}$ can be interpreted as a small `electric field'. This enables us to  write the semi-classical equation of motion for a particular band~\cite{Martin17,Crowley19}:
\begin{eqnarray}
\dot{\bm n}&=& \nabla_{\bm q}  \epsilon_{\bm q} + {\bm E} \times {\bm \Omega} \nonumber \\
\dot{\bm q}&=& {\bm E} 
\label{eq:seom}
\end{eqnarray}
where ${\bm \Omega}$ denotes the Berry curvature of the band under consideration; we consider a charge of unit magnitude for simplicity. 
In the present context, vectorized frequency is equivalent to the electric field i.e., ${\bm E} \equiv {\bm \omega}$ with ${\bm \omega}=(\omega_1, \omega_2)$.  Importantly, velocity receives an anomalous contribution from this electric field, and the direction is perpendicular to the electric field. This anomalous contribution caused by the Berry curvature ${\bm \Omega}$ plays a non-trivial role in the energy pumping dynamics.  The photon 
absorption and emission rates depend on the band dispersion and the  anomalous velocity, that has a topological origin. 

One can compute the current operator for Floquet lattice by taking the derivative of the $H_{\rm X}^{\rm IC}$ with respect to the offset phase ${J}_i=\partial H_{\rm X}^{\rm IC}/\partial \phi_i$ with $X=$ QSHI and HOTI where $\Delta=0$ and $\ne 0$, respectively.  Now, adapting the conjugate variables such as number operator ${n}_i$ and time $t$ corresponding to phase $\phi_i$ and Hamiltonian $H_{\rm X}^{\rm IC}$, respectively, the current operator reduces to the time derivative of the number operator ${J}_i= \partial \hat{n}_i/\partial t$. Physically, the rate of photon flow is captured by the operator ${\bm J}=(J_1,J_2)$ associated with two drives of frequencies ${\bm \omega}=(\omega_1,\omega_2)$. 
Interestingly, from the semi-classical equation of motion  Eq. (\ref{eq:seom}), the anomalous contribution of the $\dot{\bm n}$ is given  by ${\bm E} \times {\bm \Omega}= {\bm \omega} \times {\bm \Omega}$. The flow of photons causes the exchange of energy between the drives following the relation ${\mathcal E}_i= n_i\omega_i$. One can relate the   movement of
the particle within a band to the absorption (emission) of photon with one (other) frequency. Although the total work done by the two drives is zero as they cancel each other, individually they are finite due to photon flow.

The current operator ${J}_i$ can then further be mapped onto the rate of change of energy $d\mathcal E_i/dt$,
caused by the absorption or emission of photons with frequency $\omega_i$  as follows $\partial  {\mathcal E}_i /\partial t= \omega_i \partial {n}_i /\partial t=  \varepsilon_{ijk} \omega_i \omega_j  {\bm \Omega}_k $ where $i,j$ and $k$ are mutually orthogonal to each other. 
Note that the anomalous part of the velocity in the semi-classical equation of motion Eq. (\ref{eq:seom})  takes care of the energy exchange. 
Remarkably, the Levi-Civita symbol  $\varepsilon_{ijk}$ guarantees $\partial {\mathcal E}_i /\partial t= -\partial {\mathcal E}_j /\partial t$, indicating the fact that rate of absorption and emission are exactly opposite of each other.
 {The energy exchange in the temporal topological phase can be connected to the chirality of the edge modes in the underlying tight-binding model where $\pm k$ determines directionality. More precisely,
the momentum $k$ and $-k$ can be connected to the positive and negative slopes of work done. In particular, $ k\equiv  {\mathcal E}_1 t$ and $-k \equiv  {\mathcal E}_2 t$ which indicates the fact that energy absorption and emission, corresponding to two drive channels, are inherently connected to two opposite momenta.  In other words, $k \equiv \omega_{1,2}$  and $-k \equiv \omega_{2,1}$. 
In the case of  spin-full topological model, the momentum of edge movers is coupled with the spins such that up (down) spins move with positive (negative) chirality. This spin-momentum locking is manifested 
through the multiple distinct absorption and emission profiles for a given drive channel in the four-band temporal topological phase, as discussed in Sec. \ref{s3}.  }

\subsection*{Observables}
\label{s2ss3}

Having formulated the framework, we now derive the expression of energy exchange between the drives. In order to do that, one first needs to compute the rate of change of energy with time  $\partial {\mathcal E}_i /\partial t$ in terms of the $H_{\bm q}$ (Eq.(\ref{eq:ham_pesudoq})). This can be done by computing $\partial  {\mathcal E}_i /\partial t= \omega_i  
\big\langle \frac{\partial H_{\rm X}^{\rm IC}}{\partial \phi_i} \big\rangle= \omega_i \big\langle \frac{\partial {\bm H_{q_i}}}{\partial \phi_i}.  {\bm T} \big \rangle=  \big\langle \frac{\partial {\bm H_{q_i}}}{\partial t}.  {\bm T} \big \rangle$ where $d\phi_i \equiv \omega_i dt$, where $\langle \ldots \rangle$ can be obtained from the instantaneous time-evolved state $|\psi(t)\rangle$. For an initial state $|\psi(0)\rangle$, the time-evolved state is found to be $|\psi(t)\rangle=U(t)|\psi(0)\rangle$ with $U(t)= \mathcal{T} (\Pi_{0}^{t} e^{-i {\bm H}_q(t) dt})$ where $\mathcal{T}$ represents the time-ordered product.  The $\partial {\mathcal E}_i $ quantifies the expectation of the differential work operator $\partial W_i$ while the work operator to take the following form 
\begin{eqnarray}
    {W}_i &=& \int_{0}^{t} dt \, U(t)^\dagger \omega_i \bigg(\frac{\partial {\bm H_{q_i}}}{\partial \phi_i}.  {\bm T}\bigg) U(t) \nonumber \\
    &=&
    \int_{0}^{t} dt \, U(t)^\dagger \bigg( \frac{\partial {\bm H}_{q_i}}{\partial t}. {\bm T}\bigg) U(t).
\label{eq:work_op}    
\end{eqnarray}
As a result, the expectation value of the work is given by 
\begin{equation}
    {\mathcal E}^n_i(\Delta)=\langle W^n_i (\Delta) \rangle= \langle \psi_n (0) | {W}_i | \psi_n (0) \rangle.
\label{eq:work_exp}
\end{equation}
 {where $n=1,2,3,4$ denote the band indices of the eigenstates, arranged in ascending order, corresponding to  with the initial Hamiltonian ${\bm H_{q}}$ at $t=0$ (Eq. (\ref{eq:ham_pesudoq}))}.  Note that $\frac{\partial {\bm H}_{q_{1,2}}}{\partial t}. {\bm T}=T_{1,2} \omega_{1,2} \cos(\omega_{1,2} t + \phi_{1,2}) - T_3 \omega_{1,2} \sin(\omega_{1,2} t + \phi_{1,2}) \mp T_4 \Delta \omega_{1,2} \sin(\omega_{1,2} t + \phi_{1,2})$.  { To be precise,  the initial Hamiltonian ${\bm H_{q}}$ at $t=0$ is given by $H_{(\phi_1,\phi_2)}(\Delta)=T_1 \sin \phi_1 + T_2 \sin \phi_2  -  T_3 \big(m - \cos \phi_1 - \cos \phi_2 \big)  + \Delta T_4\big ( \cos \phi_1 - \cos \phi_2 \big)$ as obtained from Eq. (\ref{eq:ham_pesudoq}) while ${\bm q}=(q_1,q_2)=(\omega_1 t+\phi_1, \omega_2 t+\phi_2)=(\phi_1,\phi_2)$ at $t=0$. Therefore, the initial Hamiltonians are given by   $H_{\rm QSHI}^{\rm IC} (t=0)= H_{(\phi_1,\phi_2)}(\Delta=0)$ and $H_{\rm HOTI}^{\rm IC} (t=0)= H_{(\phi_1,\phi_2)}(\Delta\ne 0)$.} 
Therefore, the work done $\langle W^n_i (\Delta) \rangle$ identifies the energy emitted  (absorbed) by one (another) drive for this incommensurate drive protocol. Importantly,  $\langle W^n_i (\Delta = 0) \rangle$ ($\langle W^n_i (\Delta \ne 0) \rangle$) quantifies the energy pumping associated with the incommensurate drive for the temporal QSHI (HOTI) models.


In addition to the work done associated with the drives, we study the fidelity of the time evolved state $|\psi_n(t) \rangle$ with the  same-band instantaneous eigenstate of $H_{\bm q}(t)$ i.e., $|\tilde{\psi}_n(t) \rangle$:
\begin{equation}
    F_n(t) = |\langle \psi_n(t)|\tilde{\psi}_n(t) \rangle|^2,
\label{eq:fidelity}
\end{equation}
where \( \ket{\psi_n(t)} = U(t) \ket{\psi_n(0)}\) and the initial state, associated with the $n$-th band is $\ket{\psi_n(0)}$. In terms of the projections $P_n(t)=\big(1-{\bm H}_q(t)/\epsilon_{n,q}(t)\big)/2$,  $|\tilde{\psi}_n(t) \rangle= P_n(t) |\psi_n(t) \rangle $, the fidelity is found to be \cite{Martin17}
\begin{equation}
    F_n(\Delta,t) =  \langle \psi_n(t)|P_n(t)|\psi_n(t) \rangle= \text{Tr} \bigg | P_n(0) U(t)^{\dagger} P_n(t)  U(t)
    \bigg |
\label{eq:fidelity2}    
\end{equation}
where $\epsilon_{n,q}(t)$ is the $n$-th instantaneous eigenstate associated with Hamiltonian ${\bm H}_q(t)$ as given in Eq. (\ref{eq:ham_pesudoq}). The fidelity for the temporal QSHI and HOTI model can be computed by setting $\Delta=0$ and $\ne 0$, respectively.  Note that $F_n(t)$ is bounded between $0$ and $1$ as expected. 

Having set up the formalism for computing the energy pumping rates and the fidelity, we now discuss the framework for the computation of the topological invariant. As demonstrated earlier, one can characterize the FOTI (HOTI) phases, observed in Eq. (\ref{eq:ham_pesudoq}), using the spin-Chern number  (Wannier spectrum). Considering $0<\phi_{1,2}\le 2\pi$ as the periodic variables, the topological invariant can 
be defined, where $\phi_{1,2}$ plays the role of the lattice momentum $k_{x,y}$. Therefore, the quasi-periodically driven four-level system can be mapped to a four-band model, where $\epsilon_{\bm q}$ represents the energy associated with pseudo-momentum ${\bm q}$ in the semi-classical equation of motion Eq. (\ref{eq:seom}).  It is worth mentioning the framework of Chern number using the Berry curvature, characterizing the topologically gapped quantum anomalous Hall insulator. Adopting $0<\phi_{1,2}\le 2\pi$ as the periodic variables as stated earlier, we can construct the $z$-component of Berry curvature over a torus as given by  $\Omega^z_{12,n}(\phi_1,\phi_2) = \partial_1 A_{2,n}(\phi_1,\phi_2) - \partial_2 A_{1,n}(\phi_1,\phi_2)$ with $\partial_{1,2} =\partial /\partial \phi_{1,2}$. The Abelian Berry connection  $A_{\mu,n}(\phi_1,\phi_2) = \bra{\psi_n(\phi_1,\phi_2)} \partial_\mu \ket{\psi_n(\phi_1,\phi_2)}$ with $\mu=1,2$. The Chern number for the \(n\)th band can be expressed as \cite{Fukui2005}
\begin{equation}
C_n = \frac{1}{2\pi i} \int_{0}^{2\pi} \int_{0}^{2\pi} d\phi_1 \, d\phi_2 \, \Omega^z_{12,n}(\phi_1,\phi_2),
\label{CheNumCon}
\end{equation}   
In practice, we execute Fukui's method to compute the Chern number  \cite{Fukui2005}.  {We note that the energy pumping and deflating are observed in the topological phase of a temporal quantum anomalous Hall insulator model, having two bands, with incommensurate drives. The quantized rate of such energy transfer is characterized by the Chern number \cite{Martin17}.}  The energy pumping rate between two incommensurate drives is quantized: $\partial {\mathcal E}_{1,2}/\partial t \sim \pm  \omega_1 \omega_2 C$,
where $C \approx\int d\phi_1 d\phi_2 \Omega^k_{12}$. We note that the two-level temporal half-BHZ model, $C_1=-C_2=C$, and thus  $\partial {\mathcal E}_{1}/\partial t=- \partial {\mathcal E}_{2}/\partial t$ \cite{Martin17}.

This analysis is readily extended to the context of the spin-Chern number when we analyze the temporal QSHI model, ${\bm H_{q}}(\Delta=0)$ (Eq. (\ref{eq:ham_pesudoq})). The effective spin-projected Hamiltonian can be constructed using the eigenvectors of the valence bands $|\psi_m(\phi_1,\phi_2)\rangle$, associated with ${\bm H_{q}}(\Delta=0)$
 {
\begin{eqnarray}
    &S_{nm}(\phi_1,\phi_2) = \langle \psi_n(\phi_1,\phi_2) |  \sigma_3  \tau_0  | \psi_m(\phi_1,\phi_2) \rangle, \nonumber \\   &{\rm with}\quad  n,m = 1,2.
\end{eqnarray}
}
After diagonalizing the above operator, one can obtain 
eigenvectors $ |\tilde{\psi}_{1,2}(\phi_1,\phi_2)\rangle$ and eigenenergies $\tilde{\epsilon}_1(\phi_1,\phi_2)=-\tilde{\epsilon}_2(\phi_1,\phi_2)$
which can also be used to compute the spin-Chern numbers.  The spin-up (spin-down) Chern number is associated with $ |\tilde{\psi}_{1}(\phi_1,\phi_2)\rangle$ ($ |\tilde{\psi}_{2}(\phi_1,\phi_2)\rangle$) while a spin-gap is  always maintained $\tilde{\epsilon}_{1,2}(\phi_1,\phi_2) \ne 0$.
One can then compute the spin-Berry curvature  $\tilde{\Omega}^z_{12,n}(\phi_1,\phi_2) = \partial_1 \tilde{A}_{2,n}(\phi_1,\phi_2) - \partial_2 \tilde{A}_{1,n}(\phi_1,\phi_2)$ with $\tilde{A}_{\mu,n}(\phi_1,\phi_2) = \bra{\tilde{\psi}_n(\phi_1,\phi_2)} \partial_\mu \ket{\tilde{\psi}_n(\phi_1,\phi_2)}$ with $\mu,n=1,2$. We can adopt the Fukui method to compute the spin-Chern number as given by \cite{Fukui2005}
\begin{equation}
C_{\alpha} = \frac{1}{2\pi i} \int_{0}^{2\pi} \int_{0}^{2\pi} d\phi_1 \, d\phi_2 \, \Omega^z_{12,\alpha'}(\phi_1,\phi_2),
\label{eq:spin-Chern}
\end{equation}   
with $\alpha'=1$, and $2$ giving $\alpha=\uparrow$, and $\downarrow$, respectively. The QSHI phase can be characterized by the spin-polarized Chern number $C_{\text{spin}} = \frac{C_\uparrow - C_\downarrow}{2}$ with $C_{\text{spin}}=-1$ and $1$ for $-2<m<0$ and $0<m<2$, respectively.

\begin{figure*}[t]  
    \centering
    \includegraphics[width=\textwidth]{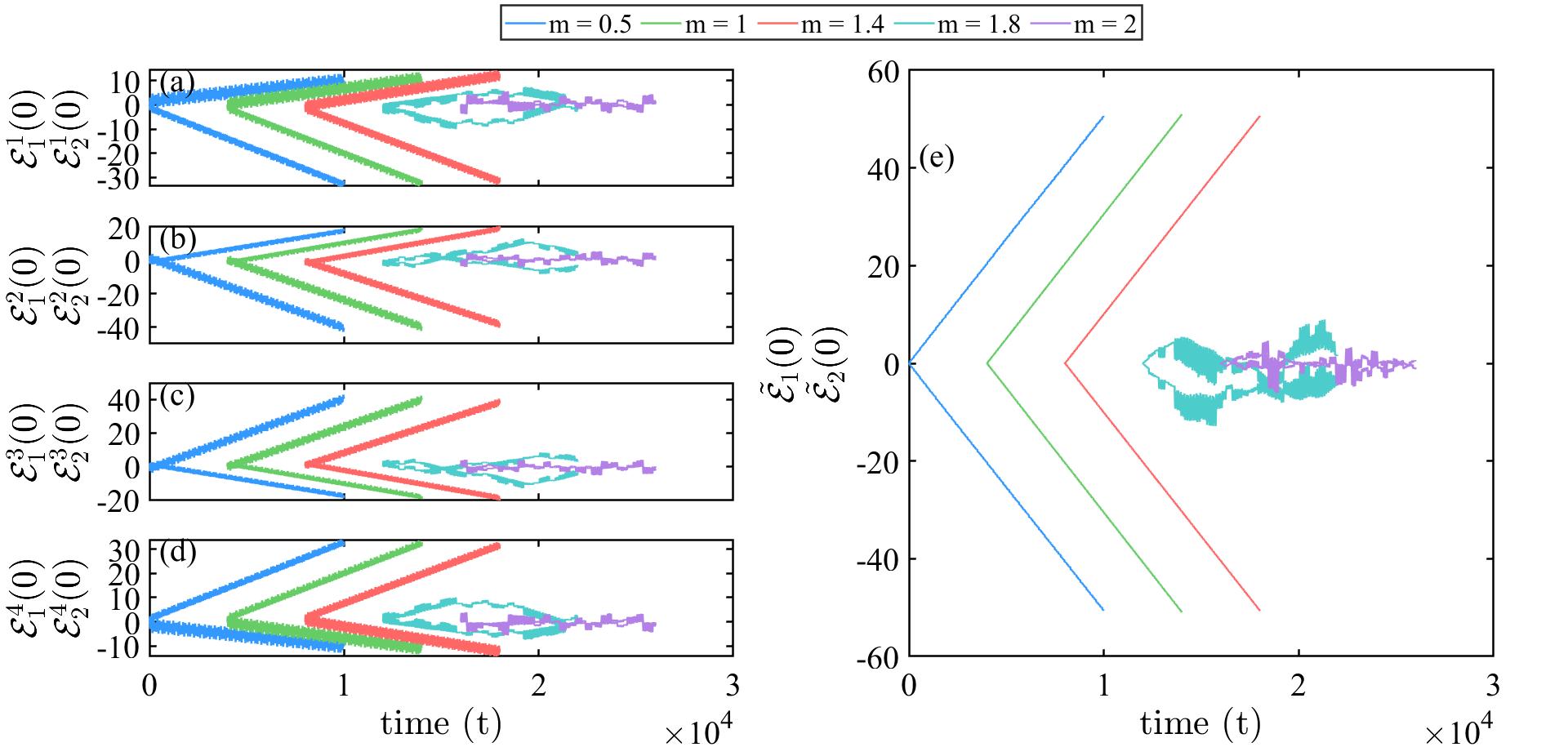} \\  
    \caption{ The work done by $i$-th drive for $n$-th band $\mathcal {E}_i^n(\Delta=0)$, computed using Eq. (\ref{eq:work_exp}), for the temporal QSHI model $H_{\rm QSHI}^{\rm IC}$,  with $i=1,2$ as a function of time for $n=1$, $2$, $3$, and $4$ is shown in (a,b,c,d) shows. The evolution of  $\mathcal {\tilde E}_i(\Delta=0)= \mathcal {E}_i^1(\Delta=0)+\mathcal {E}_i^3(\Delta=0)$ for $i=1,2$ with time is plotted in (e) where the lines with constant positive (negative) slope signify the equal rate of emission (absorption) of energy as a whole for the temporal QSHI model inside the topological phase. Note that $\mathcal {E}_i^n(0) =0$, $\tilde{\mathcal {E}}_i(0) =0$ at $t=0$ for all values of $m$ and they have been offset horizontally for clarity.
    }
    \label{fig:MainEvsTD=0}
\end{figure*}

Having demonstrated the technique to compute the FOTI topological invariant, we now discuss the HOTI topological invariant in terms of Wannier spectra. To begin with, we consider a semi-infinite geometry where the $x$-direction is open and $y$-direction periodic, represented by a good quantum number $k_y$. This analogy is carried forward to the incommensurate drive model where off-set phase $\phi_1$ ($\phi_2$) is identified with $x$ ($y$) spatial degrees of freedom. The Hamiltonian $H_{\bm q}(\Delta)$ can be written in terms of semi-infinite geometry represented by $H_{x,k_y}(\Delta)$. After diagonalizing $H_{x,k_y}(\Delta)$, we can obtain the eigenvectors $\ket{\psi_{n,\phi_1} }$ where $n$ represents the lattice index along $x$-direction and $\phi_2$ is equivalent to the momentum mode $k_y$. This exercise is done with a given value of $\omega_{1,2}$ and $t$ while $\phi_{1,2}$ constructs the  semi-infinite lattice model $H_{x,k_y}(\Delta)$ represented by a $N\times N$ Hamiltonian matrix  $H_{x,\phi_2}(\Delta)$. The occupied eigenstates $\ket{\psi_{n}(x,\phi_2) }$ with $1\le n\le N/2$, associated with $H_{x,\phi_2}(\Delta)$, are combined in a certain way to form the Wilson loop line element  $F_{mn}(x,\phi_2) = \langle \psi_{m}(x,\phi_2 + \delta \phi_2) | \psi_{n}(x,\phi_2) \rangle$ which is a  $N/2 \times N/2$ matrix and $\delta \phi_2=\frac{2\pi}{N_x}$, $N_x$ being the number of discrete points inside the Brillouin zone of $\phi_2$, $0<\phi_2\le 2 \pi$. The path-ordered product of Wilson loop line element along the periodic $\phi_2$-direction forms the Wilson loop operator for open $x$-direction 
\begin{equation}
W_{x} = F_{x,\phi_2 + (N_x - 1)\delta \phi_2} \cdots F_{x,\phi_2 + \delta \phi_2} F_{x,\phi_2}. 
\label{eq:Wlopp}
\end{equation}
Once the Wilson loop operator $W_{x,\Phi_1} $ is constructed, the Wannier Hamiltonian is defined as ${\mathcal H}_x = -i \ln W_x$. The eigenvalues of ${\mathcal H}_x$ is given by $2\pi \nu^{\rm IC}_x$ and Wannier spectrum is defined by $\nu^{\rm IC}_x \equiv {\rm mod}(\nu^{\rm IC}_x,1)$. Two isolated eigenvalues at $0.5$ of the Wannier spectrum $\nu^{\rm IC}_x$ designate the Wannier center at the mid-point of the bond, which is a key signature of the HOTI phase.  We note that the above analysis holds when $\phi_1$($\phi_2$)-direction is periodic (open) and the corresponding $0.5$ mid-gap values of $\nu^{\rm IC}_y$ identify the incommensurate drive HOTI phase.  Furthermore, in this phase, the spin-Chern number exhibits irregular behavior, thereby establishing that it is is not a legitimate topological invariant for characterizing the HOTI phase.

\section{Results}
\label{s3}

Having set up the formalism to study energy pumping and topological invariants for the temporal QSHI and HOTI phases, we now examine these quantities numerically. We work with a slightly modified version of the Hamiltonians discussed in the previous section: $\mathcal{H}^{\rm IC}_{\rm X} = H^{\rm IC}_{\rm X}/\eta$, where $\eta$ is an overall energy scale. We consider $\eta =2$, $\omega_1 = 0.1$, $\omega_2 = \omega_1(\sqrt{5}+1)/2$, time $t$ varies from $0$ to $10^4$, $\phi_1 = 0$ and $\phi_2 = \pi/10$ unless otherwise specified. We note that this choice of parameters confirms that deep in the topological phase, the band gap $\Delta = \eta \min[|m-2|,|m|]$ is large ($ \gg \omega_1, \omega_2$), thereby ensuring the validity of the semi-classical equations of motion.

\subsection*{Energy Pumping for FOTI and HOTI phases}
\label{sec:IIIA}

\begin{figure*}[t]  
    \centering
    \includegraphics[width=\textwidth]{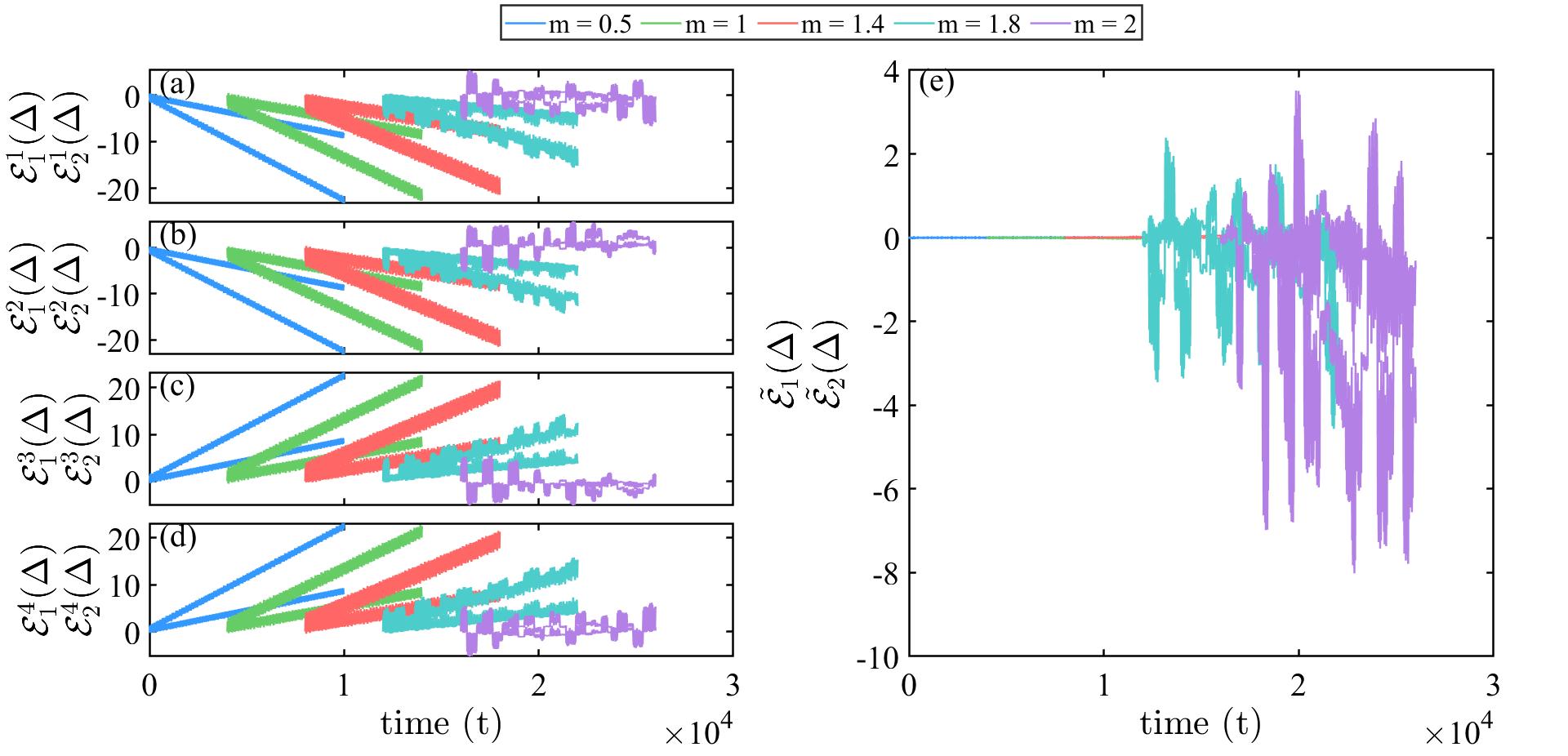}  
    \caption{ The work done by $i$-th drive for $n$-th band $\mathcal {E}_i^n(\Delta)$, computed using Eq. (\ref{eq:work_exp}), for the temporal HOTI model $H_{\rm HOTI}^{\rm IC}$, with $i=1,2$ as a function of time for $n=1$, $2$, $3$, and $4$ is shown in (a,b,c,d), respectively. The time-evolution of  $\mathcal {\tilde E}_i(\Delta)= \mathcal {E}_i^1(\Delta)+\mathcal {E}_i^3(\Delta)$ for $i=1,2$ is shown in (e) where the zero line signifies no emission or absorption of energy as a whole for HOTI model inside the topological phase. We have set $\Delta=1$ for these calculations. Note that $\mathcal {E}_i^n(\Delta) =0$, $\tilde{\mathcal {E}}_i(\Delta) =0$ at $t=0$ for all values of $m$ and they have been offset horizontally for clarity.    
    }
    \label{FInalEVsTDelta=1}
\end{figure*}

We begin our analysis by computing the work done (Eq. (\ref{eq:work_exp})) for the temporal QSHI model (Eq. (\ref{eq:ham_qshi})). The work done
${\mathcal E}^n_i(\Delta=0)$
associated with two drives $i=1,2$ and four energy-levels $n=1,2,3,4$ which are plotted in the Fig. \ref{fig:MainEvsTD=0} (a,b,c,d), respectively. To show the topological behavior, we consider $m=0.5$, $1$, $1.4$, $1.8$ and $2$, out of which work done varies  monotonically with time deep within the topological phase ($m \le 1.8$). When the phase transition is approached ($m=2$), the time-evolution of the work done gets drastically  modified leading to irregular behavior; this is a consequence of the breakdown of the quasi-adiabaticity of the dynamics. Notably, we find that ${\mathcal E}^1_i(0)=-{\mathcal E}^4_i(0)$,  ${\mathcal E}^2_i(0)=-{\mathcal E}^3_i(0)$ and  $\sum_i{\mathcal E}^{1,2}_i(0)=-\sum_i{\mathcal E}^{4,3}_i(0)$, thereby implying that  $\sum_{n}^{1,4} {\mathcal E}^n_i(0)=\sum_{n}^{2,3} {\mathcal E}^n_i(0)=0$; this is related to the chiral and particle-hole symmetry of the tight-binding QSHI model. Interestingly, ${\mathcal E}^1_i(0)$ and ${\mathcal E}^2_i(0)$ (${\mathcal E}^3_i(0)$ and ${\mathcal E}^4_i(0)$) are qualitatively similar due to the fact that bulk valence and conduction bands of the  QSHI model are doubly degenerate. The unequal magnitude of slopes for the two channels are caused by the intertwined nature of ${\bm \sigma}$ and ${\bm \tau}$ degrees of freedom in the QSHI Hamiltonian, while their sign depends on the choice of the tight-binding terms. \\

Intriguingly, we find that ${\mathcal {\tilde E}}_1(0) ={\mathcal E}^1_1(0)+ {\mathcal E}^3_1(0)$ is exactly opposite with respect to ${\mathcal {\tilde E}}_2(0)={\mathcal E}^1_2(0)+ {\mathcal E}^3_2(0)$ for all time as long as $m$ is well inside the topological region, see Fig. \ref{fig:MainEvsTD=0} (e). In other words, work done by the first (second) drive increases (decreases) linearly with time when we
consider the time evolution combining the first and third energy levels. This is a clear signature of the underlying topological feature of the model as the above characteristic is no longer observed when $m$ approaches the phase boundary $m=2$.  This feature ensures that first drive takes the energy from the second drive leading to a continuous increase (decrease) in energy for first drive  with $\omega_1$ (second drive with $\omega_2$).
We observe that this specific combination results in smooth lines, while fluctuations are observed for the behavior of individual energy levels. Importantly, the total work done remains zero when we add both the drives, which is a characteristic feature of such temporal topological models. Interestingly, ${\mathcal E}^n_1(0)$ and ${\mathcal E}^n_2(0)$ are always opposite in sign but unequal in magnitude within the topological region, irrespective of their band index. This feature is markedly distinct from the half-BHZ model where two drives are associated with exactly two opposite values for any finite time~\cite{Martin17}, and it reflects the interplay of the spin and orbital degrees of freedom in the real space QSHI model. \\

 {We now explain the results on energy pumping employing symmetry considerations. To begin with, one can consider the band dispersion $E^{n}(k)$ for $n$-th band for the tight-binding  QSHI model.  The  particle-hole symmetry $P=\sigma_3 \tau_1 K$ yields the constraint $E^1(k)=-E^3(-k)$ and $E^2(k)=-E^4(-k)$. The chiral symmetry $C= \sigma_2 \tau_1$ provides the relation $E^1(k)=-E^4(k)$ and $E^2(k)=-E^3(k)$. 
The time-reversal symmetry $T=i \sigma_2 \tau_0 K$ gives the condition $E^1(k)=E^2(-k)$ and $E^3(k)=E^4(-k)$.  We note that in the QSHI phase, $C$, $P$, and $T$ symmetries have to be satisfied simultaneously. Consequently, the relation $C= TP$ is satisfied, leading to 
$E^1(k)=-E^4(k)$ and $E^2(k)=-E^3(k)$.  In the context of the temporal QSHI model, one can find $\mathcal{E}^1_1(0)=-\mathcal{E}^4_1(0)$ and $\mathcal{E}^2_1(0)=-\mathcal{E}^3_1(0)$ as conditioned by chiral symmetry $C=PT$. This is what is observed in Fig. \ref{fig:MainEvsTD=0}. All these symmetries lead to the only  combination such that $\mathcal{E}_1^1(0)+\mathcal{E}_1^3(0)=-\mathcal{E}_2^1(0)-\mathcal{E}_2^3(0) \ne 0$ and $\mathcal{E}_1^2(0)+\mathcal{E}_1^4(0)=-\mathcal{E}_2^2(0)-\mathcal{E}_2^4(0)\ne 0$ where two drives exhibit exactly  opposite behavior. We have extensively analyzed the slope of the work done for temporal QSHI in App.~\ref{appA}. }

Having examined the QSHI model, we now investigate the work done  Eq. (\ref{eq:work_exp}) with $\Delta  \ne 0$ for the temporal HOTI model Eq. (\ref{eq:ham_hoti}) as shown in Fig. \ref{FInalEVsTDelta=1}.  We first discuss the dynamics of the individual energy levels $n=1,2,3$ and $4$ in Fig. \ref{FInalEVsTDelta=1} (a,b,c,d), respectively. The findings for HOTI model are similar to those of the QSHI model i.e.,  ${\mathcal E}^{1,2}_i(\Delta)=-{\mathcal E}^{4,3}_i(\Delta)$ and  { ${\mathcal E}^{1,3}_i(\Delta)={\mathcal E}^{2,4}_i(\Delta)$} for both the drives $i=1,2$. These outcomes are caused by the particle-hole and chiral symmetry of the underlying tight-binding model. Importantly, ${\mathcal E}^{1}_i(\Delta)+{\mathcal E}^{3}_i(\Delta)={\mathcal E}^{2}_i(\Delta)+{\mathcal E}^{4}_i(\Delta)=0$ referring to the fact that work done by the drives are zero while combining the dynamics of first and third bands; this behavior origniates from the HOT properties of the phase. Notably, unlike the previously discussed QSHI phase, for the HOTI, both the drives either release or absorb energies for a single band. However, for first (last) two bands with $n=1,2$ ($n=3,4$), the work done by both the drive is negative (positive).  Similar to the previous QSHI model,  the unequal magnitude of slopes of work done is associated with the intertwined nature of the degrees of freedom. Unlike the previous case, the identical sign of work done, associated with the two drive channels, is caused by the additional four-fold rotational symmetry breaking tight-binding term in HOTI model which is absent in QSHI or quantum anomalous Hall insulator models.\\    

 {It is instructive to further analyze the combination of bands for the temporal HOTI phase.  The same combination of bands as QSHI phase  holds in the case for the HOTI phase where slope itself vanishes, see Fig. \ref{FInalEVsTDelta=1}.  
In the case of temporal HOTI phase,  chiral symmetry $C$ ensures 
$\mathcal{E}^1_1(\Delta)=-\mathcal{E}^4_1(\Delta)$, $\mathcal{E}^1_2(\Delta)=-\mathcal{E}^4_2(\Delta)$ and $\mathcal{E}^2_1(\Delta)=-\mathcal{E}^3_1(\Delta)$, $\mathcal{E}^2_2(\Delta)=-\mathcal{E}^3_2(\Delta)$ while particle-hole symmetry $P$ conditions $\mathcal{E}^1_1(\Delta)=-\mathcal{E}^3_2(\Delta)$, 
$\mathcal{E}^1_2(\Delta)=-\mathcal{E}^3_1(\Delta)$
and $\mathcal{E}^2_1(\Delta)=-\mathcal{E}^4_2(\Delta)$, $\mathcal{E}^2_2(\Delta)=-\mathcal{E}^4_1(\Delta)$. This results in $\mathcal{E}^1_{1,2}(\Delta)=\mathcal{E}^2_{2,1}(\Delta)$ and $\mathcal{E}^3_{1,2}(\Delta)=\mathcal{E}^4_{2,1}(\Delta)$. 
As the TRS $T$ is broken in HOTI phase, $P$ and $C$ compositely lead to the same combination such that  
$\mathcal{E}_1^1(\Delta)+\mathcal{E}_1^3(\Delta)=-\mathcal{E}_2^1(\Delta)-\mathcal{E}_2^3(\Delta)=0$ and $\mathcal{E}_1^2(\Delta)+\mathcal{E}_1^4(\Delta)=-\mathcal{E}_2^2(\Delta)-\mathcal{E}_2^4(\Delta)=0$. Therefore, this zero slope corresponds to the no transfer of energy between the drives indicating towards the breakdown of quantized transport. The emission and absorption for a given band is a consequence of time-reversal symmetry which is observed only in temporal QSHI phases. Therefore, the identical signs of the slope of work done can be attributed to the breaking of time-reversal symmetry as observed in  temporal HOTI phases. We refer the reader to App.~\ref{appA} for more details.}

Next, we proceed to analyze the connection between the net energy exchange and the underlying topology of the phase. On the one hand, for the QSHI phase, the linear variation of ${\mathcal {\tilde E}}_i(0)={\mathcal E}^1_i(0)+{\mathcal E}^3_i(0)$ with exactly opposite slope represents the existence of propagating one-dimensional edge modes in the corresponding tight-binding model (see Fig. \ref{fig:MainEvsTD=0} (e)). On the other hand, the constant line of ${\mathcal {\tilde E}}_i(\Delta)={\mathcal E}^1_i(\Delta)+{\mathcal E}^3_i(\Delta)$ with zero slope indicates the presence of localized zero-dimensional  corner modes in the two-dimensional HOTI phase (see Fig. \ref{FInalEVsTDelta=1} (e)). Thus, the nature of the boundary modes is perfectly captured by the energy dynamics between the drives. The  non-topological phases do not show any such clear behavior, except for the fluctuations around $0$ for both models. Our analysis demonstrates a clear route to realize and characterize temporal topological phases in quasi-periodically driven qudit systems.  {We note that the slope of the work done depends on the parameters of the underlying tight-binding model, such as the spin-orbit coupling, hopping strength, and magnitude of Wilson-Dirac mass term. The rate of change of energy exchange between the drives varies monotonically with the parameters in the strong driving limit above a threshold value of the parameters. This is a hallmark of quantization, where  the slopes of the work done for the chiral partner bands are related to each other. We refer the reader to Apps.~\ref{appB} and \ref{appC} for more details.}  \\ 

\subsection*{Topological characterization of FOTI and HOTI phases}

\begin{figure}[b]
    \centering
    \includegraphics[width=\columnwidth]{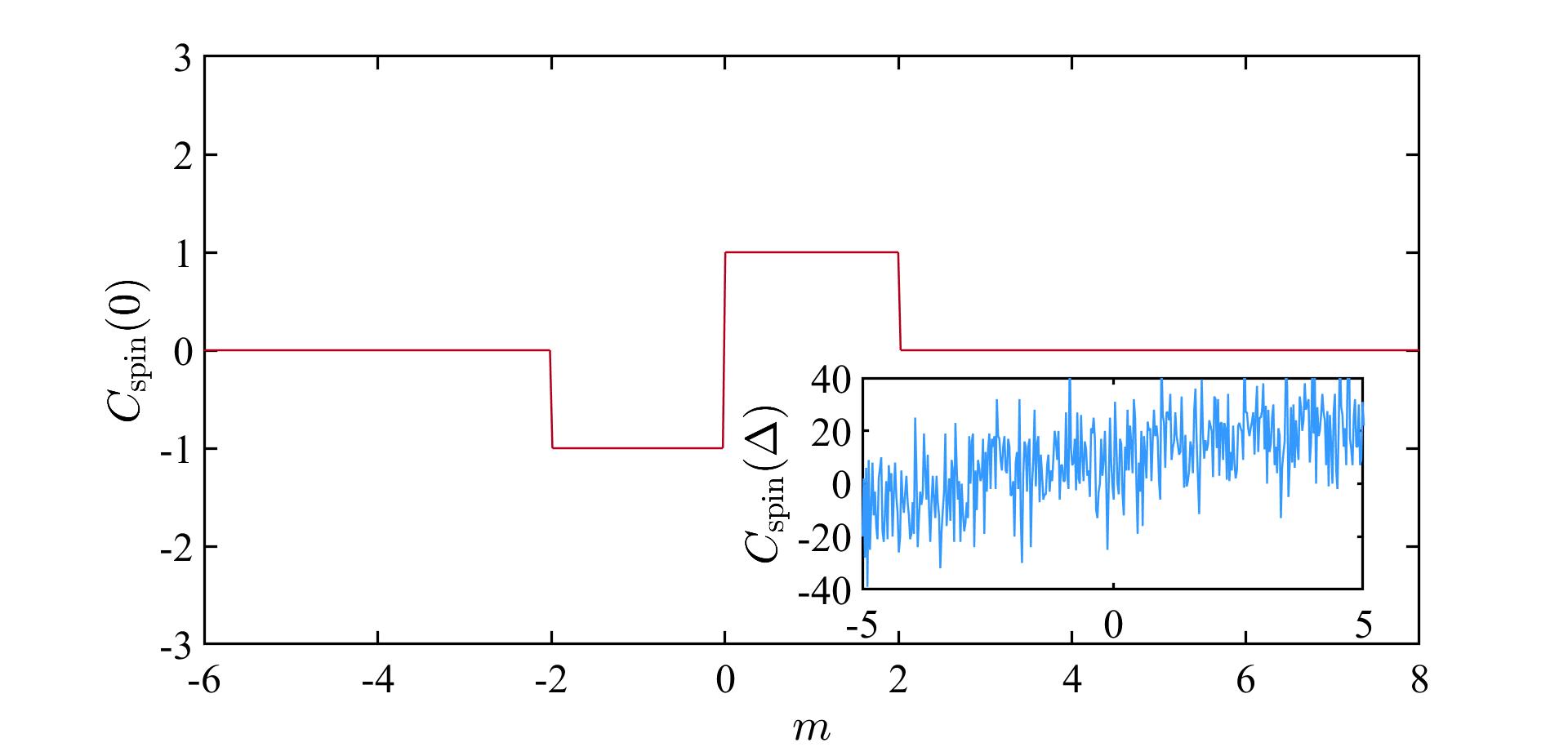}
    \caption{We depict the variation of spin-Chern Number $C_{\rm spin} (\Delta=0)$, using Eq. (\ref{eq:spin-Chern}), as a function of \( m \) for \( \Delta = 0 \). The  QSHI phases are characterized by  the quantized value of $C_{\rm spin}=\pm 1$ within $-2<m<2$ region. The inset shows the irregular 
    nature of spin-Chern Number $C_{\rm spin} (\Delta \ne 0)$ as a function of \( m \) for \( \Delta = 0.5 \), justifying the impropriety of the first-order QSHI invariant to characterize the HOTI phase.}
    \label{fig:SCDelta=0}
\end{figure}

\begin{figure}[b]
    \centering
    \includegraphics[width=\columnwidth]{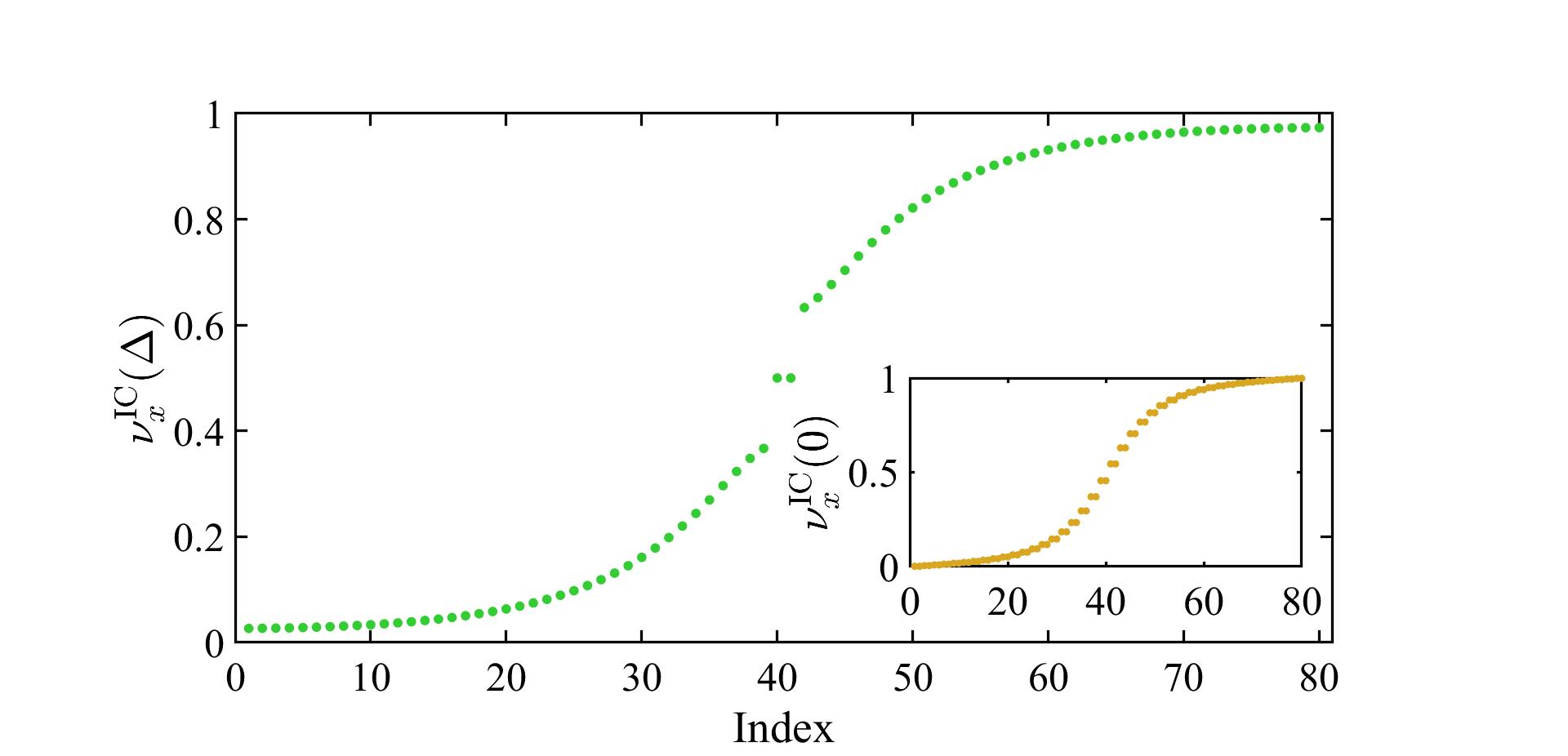}
    \caption{We show the Wannier spectrum $\nu_x^{IC} (\Delta)$, computed using Eq. (\ref{eq:Wlopp}), for \( \Delta = 0.4 \) with \( m = 1.4 \) and number of $x$-site index \( N_x = 40 \). We observe two Wannier centers at 0.5, indicating localized corner modes in the HOTI phase. In the insets, we plot the Wannier spectrum $\nu_x^{IC} (0)$ for \( \Delta = 0 \), where a continuous spectrum appears, justifying the impropriety of HOTI invariant to characterize the first-order QSHI phase.}
    \label{fig:WS=0}
\end{figure}
Given the fact that energy exchange between the drives  or work done  by the drives are connected to the topology of the underlying phase, we now examine the bulk topological invariant to strengthen the above connection. We first examine the quantized energy pumping in the QSHI phase in relation to the spin Chern number, Eq. (\ref{eq:spin-Chern}). 
We plot the spin-polarized Chern number $C_{\rm spin}$ using Eq. (\ref{eq:ham_pesudoq}) in Fig. \ref{fig:SCDelta=0}, where its quantized values qualitatively explain the constant linear slope of the work done ${\mathcal {\tilde E}}_{1,2}(0)$. Outside the topological window $|m| > 2$, $C_{\rm spin}$ vanishes, indicating the absence of quantized energy pumping. This is in complete agreement with Fig. \ref{fig:MainEvsTD=0}, where the work shows fluctuations around zero near the phase boundary $m = 2$.

 {
We now discuss the quantized rate of the  energy pumping following the spin Chern number analysis. To begin with, we note that time-reversal symmetry operator $T=i \sigma_2 \tau_0 K$ combines $n=1$ ($3$) and $2$ ($4$) bands, and therefore, $n=1,3$ and $2,4$ bands are associated with identical spin sectors. Consequently, the band indices $n=1,3$ and $2,4$ collectively yield the 
spin contributions to the dynamics. One can thus obtain $\mathcal { E}_{i}^{\uparrow}(\Delta=0)=\mathcal {E}_i^1(\Delta=0)+\mathcal {E}_i^3(\Delta=0)$ and $\mathcal { E}_{i}^{\downarrow}(\Delta=0)=\mathcal {E}_i^2(\Delta=0)+\mathcal {E}_i^4(\Delta=0)$ where subscript $i$ denotes the drive channels and superscipt refers to the band index.
Therefore, one can identify the total contribution arising from a given spin sector in terms of the following: 
$ \mathcal { E}_{\uparrow}(\Delta=0)=  \mathcal { E}^{\uparrow}_1(\Delta=0) -  \mathcal { E}^{\uparrow}_2(\Delta=0)$ and $ \mathcal { E}_{\downarrow}(\Delta=0)=  \mathcal { E}^{\downarrow}_1(\Delta=0) -  \mathcal { E}^{\downarrow}_2(\Delta=0)$. Note that $\mathcal { E}^{\uparrow}_1(\Delta=0) =- \mathcal { E}^{\downarrow}_1(\Delta=0)$ and $\mathcal { E}^{\uparrow}_2(\Delta=0) =- \mathcal { E}^{\downarrow}_2(\Delta=0)$ which ensures the preservation of time-reversal symmetry.. This can be attributed to the fact that $\uparrow$ and $\downarrow$ spin-degrees of freedom contribute to the energy pumping
separately in an exact opposite manner when the drive channels are summed up. The total rate of energy exchange between the spin sectors cancels out leading to the following profiles of work done $\mathcal { E}_{\uparrow}(\Delta=0) + \mathcal { E}_{\downarrow}(\Delta=0)=0$.}

 { It can be anticipated from the above analysis that $\mathcal { E}_{\uparrow}(\Delta=0)$ can be connected with $C_{\uparrow} $ while $\mathcal {E}_{\downarrow}(\Delta=0)$ with $C_{\downarrow}$. 
Therefore, mathematically, $\frac{d\mathcal { E}_{\uparrow}(\Delta=0)}{dt}=\frac{C_{\uparrow}}{2\pi}\omega_1\omega_2$ and $\frac{d\mathcal { E}_{\downarrow}(\Delta=0)}{dt}=\frac{C_{\downarrow}}{2\pi}\omega_1\omega_2$. The difference in pumping between the two spin sectors remains quantized as given below
\begin{equation}
    \frac{d\Delta \mathcal {E}}{dt}  = \frac{d\mathcal { E}_{\uparrow}(\Delta=0)}{dt} - \frac{d\mathcal { E}_{\downarrow}(\Delta=0)}{dt}= \frac{2C_s}{2\pi}\omega_1\omega_2 = \frac{C_s}{\pi}\omega_1\omega_2.
\end{equation}
Hence, the observation of  quantized  energy transfer in the temporal QSHI phase is governed by the spin Chern number serving as the relevant topological invariant. }

 {The above analogy holds true for QSHI phase where $\mathcal { E}_{\uparrow}(\Delta=0)$ and $\mathcal { E}_{\downarrow}(\Delta=0)$ remains finite. In the case of HOTI with $\Delta\ne 0$, $\mathcal { E}^{\uparrow}_{1,2} =\mathcal { E}^{\downarrow}_{1,2}=0$ leading to $\mathcal { E}_{\uparrow}=0$ and $\mathcal { E}_{\downarrow}=0$ 
which causes a breakdown of the quantized transport. To be precise,   there is no transport at all taking place in HOTI phase.  The disruption of 
the quantized  energy transfer to no energy transfer in the temporal HOTI phase is a direct consequence 
of the breakdown of spin Chern number. We 
compute  $C_{\rm spin}$ in the HOTI phase, as shown in the inset of Fig. \ref{fig:SCDelta=0}, where the fluctuation clearly signifies the breakdown of the first-order topological invariant. }

To circumvent this issue, we compute the Wannier spectra $\nu_x^{\rm IC}$ using Eq. (\ref{eq:Wlopp}) for the HOTI phase as shown in Fig. \ref{fig:WS=0} \cite{Ghosh21,Ghosh21b,schindler2018higher,Ghosh21c}. We find two mid-gap Wannier spectra $\nu_x^{\rm IC}=0.5$ signifying the existence of the corner modes   
in HOTI phase. This suggests the vanishing energy exchange between the drives. We also cross-check the continuous Wannier spectra $\nu_x^{\rm IC}$ without any mig-gap eigenvalues with $\Delta=0$  signifying the  first-order topological phase, shown as insets in Fig. \ref{fig:WS=0}. We conclude that this bulk invariant analysis provides a powerful tool to characterize the energy exchange between the drives.  This behavior clearly justifies the presence (absence) of 
the energy pumping for $\Delta =0$ (finite $\Delta$) in the temporal QSHI (HOTI) phases and which is again consistent with the boundary transport characteristics of the spatial QSHI (HOTI) phases.

\subsection*{Fidelity}

\begin{figure}[htbp!]
    \centering
    \includegraphics[width=\columnwidth]{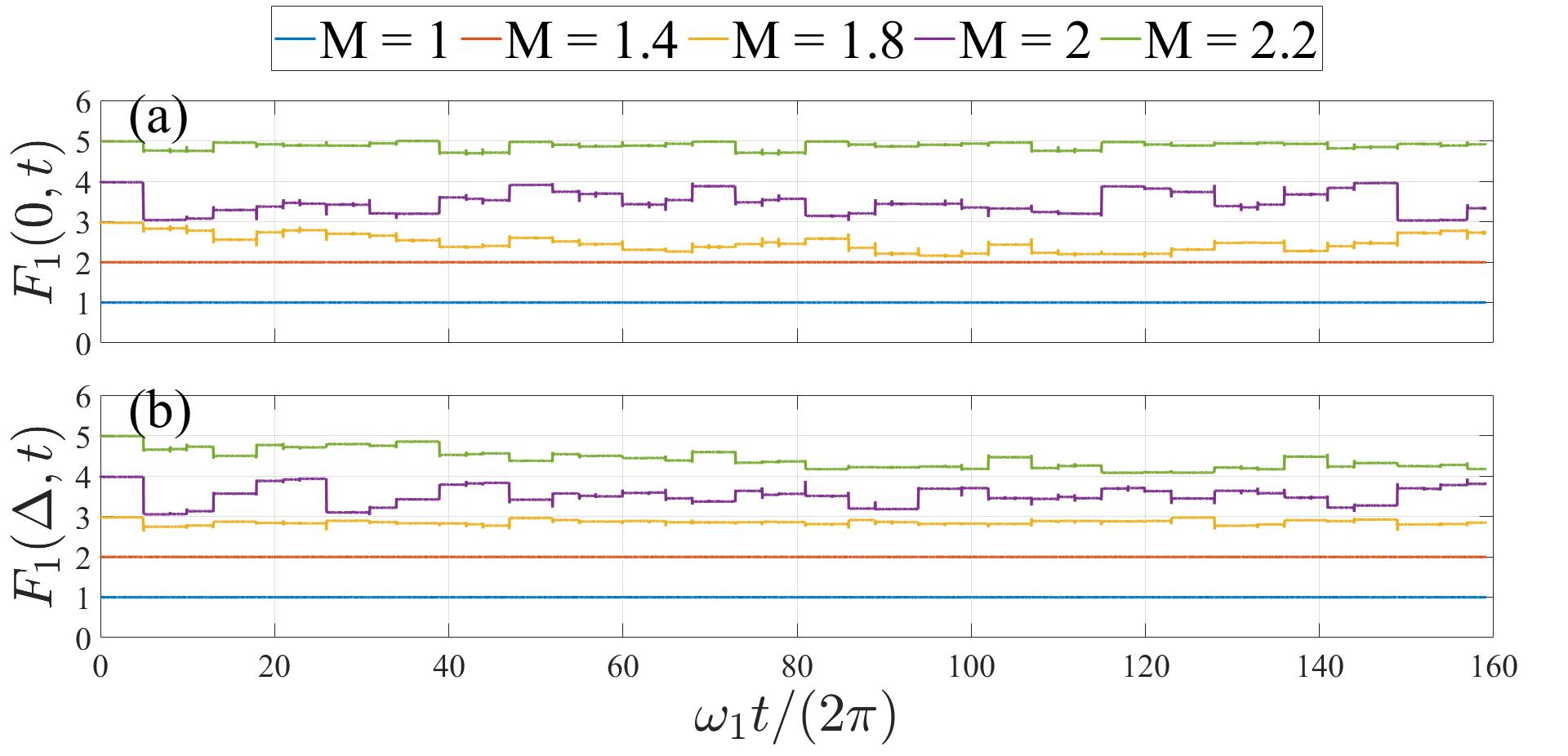}
    \caption{We show the evolution of fidelity $F_n(\Delta,t)$, computed using Eq. (\ref{eq:fidelity2}), for QSHI Eq. (\ref{eq:ham_qshi}) and HOTI  Eq. (\ref{eq:ham_hoti}) models in (a,b), respectively, with $\Delta=0$ and $\Delta =1$. We observe perfect and imperfect behavior of fidelity is connected with the topological and trivial nature of the underlying model respectively.  We consider $n=1$ for our analysis. $F_1 (t=0) = 1$ for all values of $m$ and it has been offset vertically for clarity.}
    \label{FidelityDelta}
\end{figure}

We round out our discussion of the temporal QSHI and HOTI models  by examining the fidelity $F_n(\Delta,t)$ (Eq. (\ref{eq:fidelity2})) between the instantaneous $n$-th eigenstate and the corresponding time-evolved states. The results for  $F_1(\Delta,t)$ with $\Delta = 0$ and $\ne 0$ for QSHI and HOTI models in Fig. \ref{FidelityDelta} (a,b), respectively. We note that for both models, the fidelity remains constant without any fluctuations throughout the evolution deep inside the topological phase. However, we observe severe fluctuations when $m$ approaches the phase boundary, which is seen for $m\ge 1.8$. Therefore, the constant (perfect) and fluctuating (imperfect) behavior mark the topological phase (phase transition). We qualitatively observe the same behavior for other band indices as well. The constant nature of the fidelity in the topological phase can be attributed to the quasi-adiabatic nature of the dynamics, when the driving frequencies, $\{\omega_1, \omega_2\} \ll \Delta$, where $\Delta$ is the band-gap; qualitatively similar behavior has been observed for the temporal half-BHZ model~\cite{Martin17}. The perfect nature of the fidelity results in regular and quantized energy exchange between the drives. We note however that when the system approaches the topological phase transition ($m \sim 2$), the quasi-adiabatic nature of the dynamics breaks down, resulting in strong fluctuations in $F_n(\Delta,t)$; these flucatuations are reflected in the energy exchange dynamics. We conclude that the time-evolution of the fidelity serves as a signature of the topological phase and phase transition. 

\section{Summary and Outlook}
\label{s4}

In this work, we have explored the temporal topological properties of quasi-periodically driven four-level systems that can be mapped to four-band QSHI and HOTI models in frequency space. Our primary focus was to study the energy pumping between the drives in these systems and explore its connection to the underlying topology of the corresponding tight-binding models.  More precisely, the momentum degrees of freedom in a tight-binding model are disguised to host the Floquet lattice in the Fourier space of photon sector and
the off-set phase can be considered as a periodic variable
to capture the bulk topology of the Fourier space lattice.

To begin with, we have investigated the temporal QSHI model whose real-space counterpart is characterized by intertwined spin and orbital degrees of freedom. This system exhibits a richer structure in energy flow compared to the quasi-periodically driven qubit that has been studied in earlier works~\cite{Martin17,crowley2020half}. We have computed the energy exchange between the drives, quantified in terms of work done, and found that the energy absorption and emission occurs with constant but different rates for a given band index. These rates for different bands are connected by the underlying chiral symmetry of the model. Consequently, the contributions after summing over the chiral partners yield constant but exactly opposite rates of emission (absorption) of energy by one (other) drive. This feature is a direct signature of the first-order topology in this system, where helical edge modes propagate along the one-dimensional edges in a two-dimensional QSHI. We further validate this conclusion from the quantized value of the spin-Chern number, thereby establishing a correspondence between the bulk topological invariant and the quantized rate of energy exchange.   { The energy dynamics of the bands in temporal quantum spin Hall insulator phases  are governed by chiral, particle-hole, and time-reversal symmetries.}

We perform a similar analysis on the temporal HOTI model where the higher-order mass term yields a zero exchange of energy, obtained after summing over the chiral partners, between the drives due to the presence of localized corner modes. Intriguingly, unlike both the QSHI and the half-BHZ model, for each individual band, both the drives either absorb or emit energy. We further characterize this phase by computing the second-order invariant, namely the Wannier spectra, and demonstrate that it hosts half-integer mid-gap modes which is a salient feature of the HOTI phase.  { The energy dynamics of the bands in temporal higher-order topological insulator phases are governed by chiral, particle-hole symmetries.} Finally, we examine the fidelity of the time-evolved states, associated with a given band, and demonstrate that the fidelity for a given band remains constant ($\sim 1$) deep in the topological phase; this leads to regular and quantized energy pumping. However, the fidelity fluctuates strongly when the phase transition is approached, thereby resulting in irregular energy pumping. Finally, we note that our predictions can be verified in various experimental platforms such as superconducting quantum processors~\cite{liu2023performing,tripathi2025qudit,hrmo2023native}, Rydberg atom arrays~\cite{mitra2023qudit,gonzalez2022hardware}, trapped ion crystals~\cite{ringbauer2022universal,low2025control}, ultracold atoms in optical lattices~\cite{flament2025unitary}, Nitrogen vacancy centers in diamond~\cite{zhou2024robust}, and photonic systems~\cite{dong2023highly,wang2023photonic}, where qudits can be realized and controlled yielding the possible experimental realizations of topological edge modes \cite{wang2009observation,Xiao-Dong19,Zhaoju15,he2016acoustic,jotzu2014experimental,lohse2016thouless,Hei24}.

There are several avenues for future studies following up on this work. Firstly, it would be interesting to investigate the topological phases that can be generated in a qudit driven by a $D-$tone quasi-periodic protocol with $D>2$. The fate of energy pumping in quasiperiodically driven qudit models with disorder can be another interesting direction of work. Finally, it would be intriguing to investigate heating and quantum information dynamics in quasi-periodically driven many-body qudit  systems.

\section{Acknowledgements}
T.N. acknowledges the NFSG “NFSG/HYD/2023/H0911” from BITS Pilani. S.C. thanks DST, India, for support through the project DST/FFT/NQM/QSM/2024/3.T.N. and thanks the Advanced Research Grant (ARG) from Anusandhan National Research Foundation Grant No. ANRF/ARG/2025/003163/PS.

\appendix
\section{ {Slope analysis of energy pumping from chiral partner bands}}
\label{appA}

\subsection{Temporal QSHI Phase}

 {
We first analyze the temporal quantum spin Hall insulator (QSHI) phase. We below show the numerical analysis of the slopes associated with the combinations. 
Physically, this combination captures the net energy transferred between the two drives, filtering out local oscillations which are may be an artifact of the intra-band transitions.  
To quantify this, we fitted the time evolution of the work done $\mathcal {E}_i^{n}(t)$ to extract their slopes, which represent the average pumping rates. The extracted slopes are as follows:
\[
\begin{aligned}
&\text{Initial state,}~~  n=1:  \frac{d\mathcal{E}_1^1}{dt} = -0.003112, \quad \frac{d\mathcal{E}_2^1}{dt} = 0.001065, \\
&\text{Initial state,}~~  n=2: \frac{d\mathcal{E}_1^2}{dt} = 0.001983, \quad \frac{d\mathcal{E}_2^2}{dt} = -0.004029, \\
&\text{Initial state,}~~  n=3: \frac{d\mathcal{E}_1^3}{dt} = -0.001983, \quad \frac{d\mathcal{E}_2^3}{dt} = 0.004029, \\
&\text{Initial state,}~~  n=4: \frac{d\mathcal{E}_1^4}{dt} = 0.003112, \quad \frac{d\mathcal{E}_2^4}{dt} = -0.001065.
\end{aligned}
\]
We note that the slopes associated with $n=1$  and $4$  appear in pairs of opposite signs for the drives. The same observation holds true for $n=2$ and $3$. This is in accordance with the symmetry argument as discussed earlier. The pumping contributions are exactly opposite for two drive channels when the $n=1$ and $3$ bands are summed. To be precise, 
\[
\frac{d}{dt}(\mathcal{E}_1^1 + \mathcal{E}_1^{3}) = -0.005095, \quad
\frac{d}{dt}(\mathcal{E}_2^{1} + \mathcal{E}_2^{3}) = 0.005095,
\]
we observe that their magnitudes are equal and opposite, leading to
\[
\frac{d}{dt}\big[(\mathcal{E}_1^{1} + \mathcal{E}_1^{3}) + (\mathcal{E}_2^{1} + \mathcal{E}_2^{3})\big] = 0.
\]
This exact cancellation demonstrates a quantized balance of energy flow between the two drives—i.e., the total pumped energy per cycle is topologically constrained to zero, while each channel exhibits a linear (quantized) rate of emission or absorption.
The same analysis applies to $n=2$ and $4$ finally yielding 
\[
\frac{d}{dt}\big[(\mathcal{E}_1^{2} + \mathcal{E}_1^{4}) + (\mathcal{E}_2^{2} + \mathcal{E}_2^{4})\big] = 0.
\]
}

\subsection{Temporal HOTI Phase}

 {
We extend the  analysis on  the temporal HOTI phase where $\Delta \neq 0$. Similar to the temporal QSHI case, we perform a numerical analysis of the slopes associated with the work done by the two drives. As before, we focus on the combinations of chiral partner bands, which capture the net energy exchange between the drives while suppressing local oscillations arising from intra-band processes. The slopes extracted from the time evolution of $\mathcal{E}_i^{n}(t)$ are given below:
\[
\begin{aligned}
&\text{Initial state,}~~ n=1:  \frac{d\mathcal{E}_1^1}{dt} = -0.000763, \quad \frac{d\mathcal{E}_2^1}{dt} = -0.002006, \\
&\text{Initial state,}~~ n=2:  \frac{d\mathcal{E}_1^2}{dt} = -0.000765, \quad \frac{d\mathcal{E}_2^2}{dt} = -0.001998, \\
&\text{Initial state,}~~ n=3:  \frac{d\mathcal{E}_1^3}{dt} = 0.000764, \quad \frac{d\mathcal{E}_2^3}{dt} = 0.002002, \\
&\text{Initial state,}~~ n=4:  \frac{d\mathcal{E}_1^4}{dt} = 0.000763, \quad \frac{d\mathcal{E}_2^4}{dt} = 0.002002 .
\end{aligned}
\]
We observe that the slopes for the first two bands ($n=1,2$) are negative for both drives, while those for the last two bands ($n=3,4$) are positive. Thus, the lower bands release energy whereas the higher bands absorb energy. As discussed in the main text, this structure is consistent with the symmetry properties of the HOTI Hamiltonian.
We now examine the combination of the first and third bands. Summing up the associated contributions, we obtain 
\[
\frac{d}{dt}(\mathcal{E}_1^1 + \mathcal{E}_1^{3}) = 0.000001, \quad
\frac{d}{dt}(\mathcal{E}_2^{1} + \mathcal{E}_2^{3}) = -0.000004,
\]
which are essentially zero within numerical precision. Consequently,
\[
\frac{d}{dt}\big[(\mathcal{E}_1^{1} + \mathcal{E}_1^{3}) + (\mathcal{E}_2^{1} + \mathcal{E}_2^{3})\big] = 0 .
\]
This indicates that, unlike the QSHI phase, the combined slopes vanish and no net energy transfer occurs between the two drives. The same behavior is obtained for the combination of $n=2$ and $n=4$. Hence, the temporal HOTI phase does not exhibit a finite pumping rate between the drives, in contrast to the linear energy pumping observed in the QSHI phase.}

\section{ {The effect of  spin-orbit coupling on the slopes of energy exchange}}
\label{appB}

 {
The energy pumping contributions $E_i^{(n)}$ are indeed not strictly quantized when the strength of spin-orbit coupling $\lambda$ is varied.  This behavior originates from the intertwined nature of the spin (${\bm \sigma}$) and orbital (${\bm \tau}$) sectors in the QSHI Hamiltonian, which we write as
\begin{equation}
    H^{\rm IC}_{\rm QSHI} = \eta \left[ \sum_{i=1}^{2} \lambda \, T_i S_i - T_3 \bigg( m - \sum_{i=1}^{2} C_i \bigg) \right],
    \label{eq:ham_qshi1}
\end{equation}
where \(\lambda\) denotes the spin-orbit coupling strength, and $\eta$ labels the energy scale.}

\begin{figure}[t]
    \centering
    \includegraphics[width=0.9\linewidth]{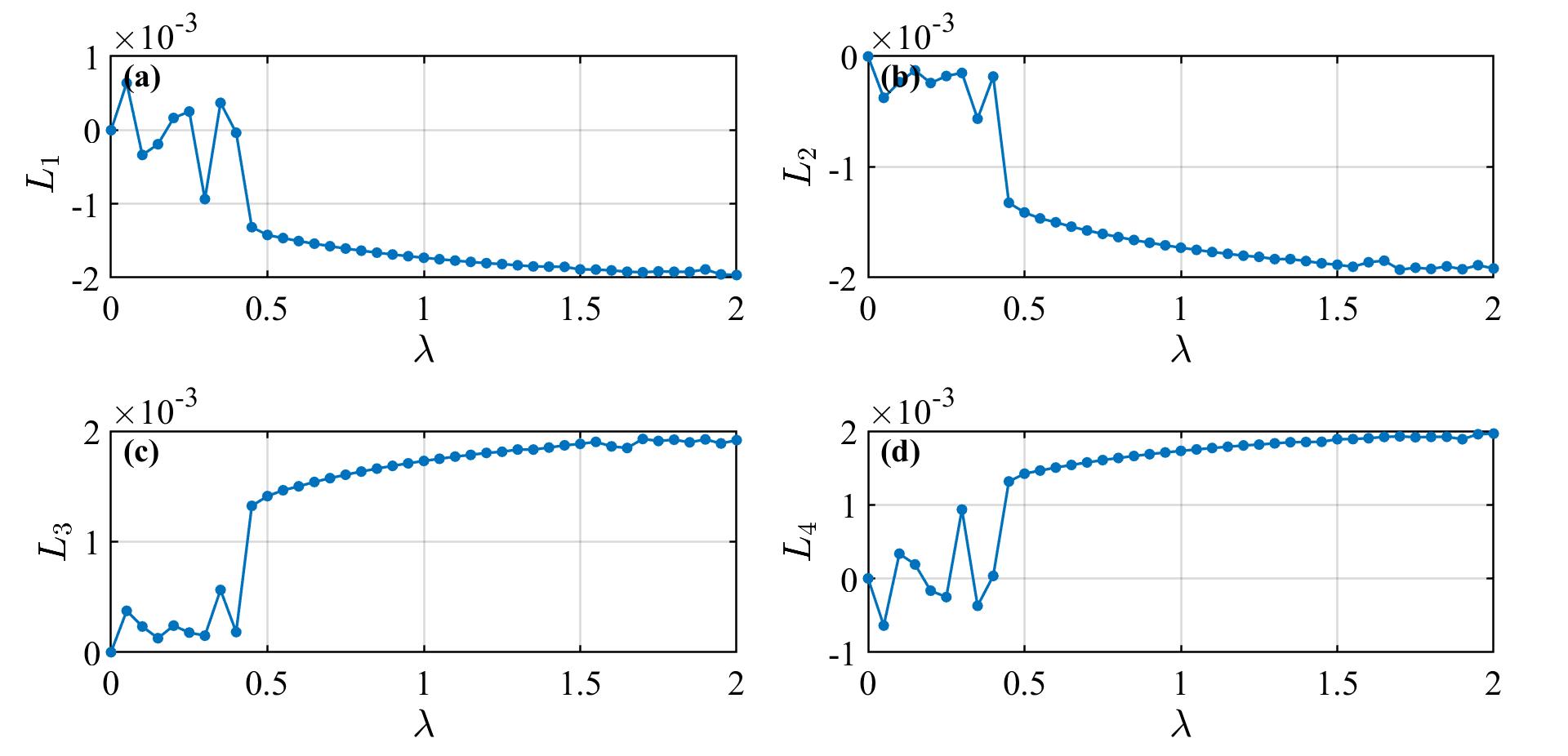}
    \caption{ {We show the behavior of the slopes $L_n =\frac{d\mathcal{E}^n_{\rm sum}}{dt}=\frac{d\mathcal{E}_1^{n}}{dt}+ \frac{\mathcal{E}_2^{n}}{dt}$ for $m = 1.4$ with the strength of spin-orbit coupling $\lambda$ for $n=1,2,3$ and $4$ in (a,b,c) and (d), respectively. It can be observed that   $L_{1,2}=-L_{4,3}$. }
    }
    \label{fig:E1E2_slope_vs_a}
\end{figure}

 {
We compute $\mathcal{E}^n_{\rm sum}=\mathcal{E}_1^{n}+ \mathcal{E}_2^{n}$ after summing over the drive channels for band $n=1,2,3$, and $4$ in the QSHI phase. Note that $\mathcal{E}_1^{n}$ and $\mathcal{E}_2^{n}$ are not opposite to each other and hence their sum can be a good indicator to investigate the effect of spin-orbit coupling. We compute the slopes  $L_1=\frac{d\mathcal{E}^1_{\rm sum}}{dt}$, $L_2=\frac{d\mathcal{E}^2_{\rm sum}}{dt}$, $L_3=\frac{d\mathcal{E}^3_{\rm sum}}{dt}$ and $L_4=\frac{d\mathcal{E}^4_{\rm sum}}{dt}$ as a function of $\lambda$ in Fig.  \ref{fig:E1E2_slope_vs_a} (a,b,c), and (d), respectively.  The slope fluctuates around zero (small negative and positive) value for small  $\lambda$ in $n=1,4$ ($2$ and $3$, respectively) bands. Importantly, for $\lambda < 0.5$, the system is outside the strong-driving regime, and the adiabatic condition is not satisfied. The slope shows a monotonic behavior for  $\lambda \ge 0.5$ in all the bands. The profiles of the slopes are exactly opposite for $n=1$ ($2$) and $4$ ($3$) for any finite value of $\lambda$. The monotopic profile of the slopes for $\lambda > 0.5$ is a consequence of energy pumping. This  leads to the fact that energy exchanges at a quantized rate and this rate depends on the strength of spin-orbit coupling.   As $\lambda$ increases above $0.5$, the system enters the strong-driving regime, allowing the semiclassical equations of motion to hold. Here, the slopes stabilize, and linear behavior of the energy exchange emerges.  In the monotonic regime with $\lambda > 0.5$, the energy pumping takes place at a  quantized rate. The spin-orbit coupling plays an important role in determining the slope. 
The  positive (negative) values of the slopes for $n=3,4$  ($n=1,2$) are observed for $ \lambda>0.5$.    }

\begin{figure}[h!]
    \centering
    \includegraphics[width=1\linewidth]{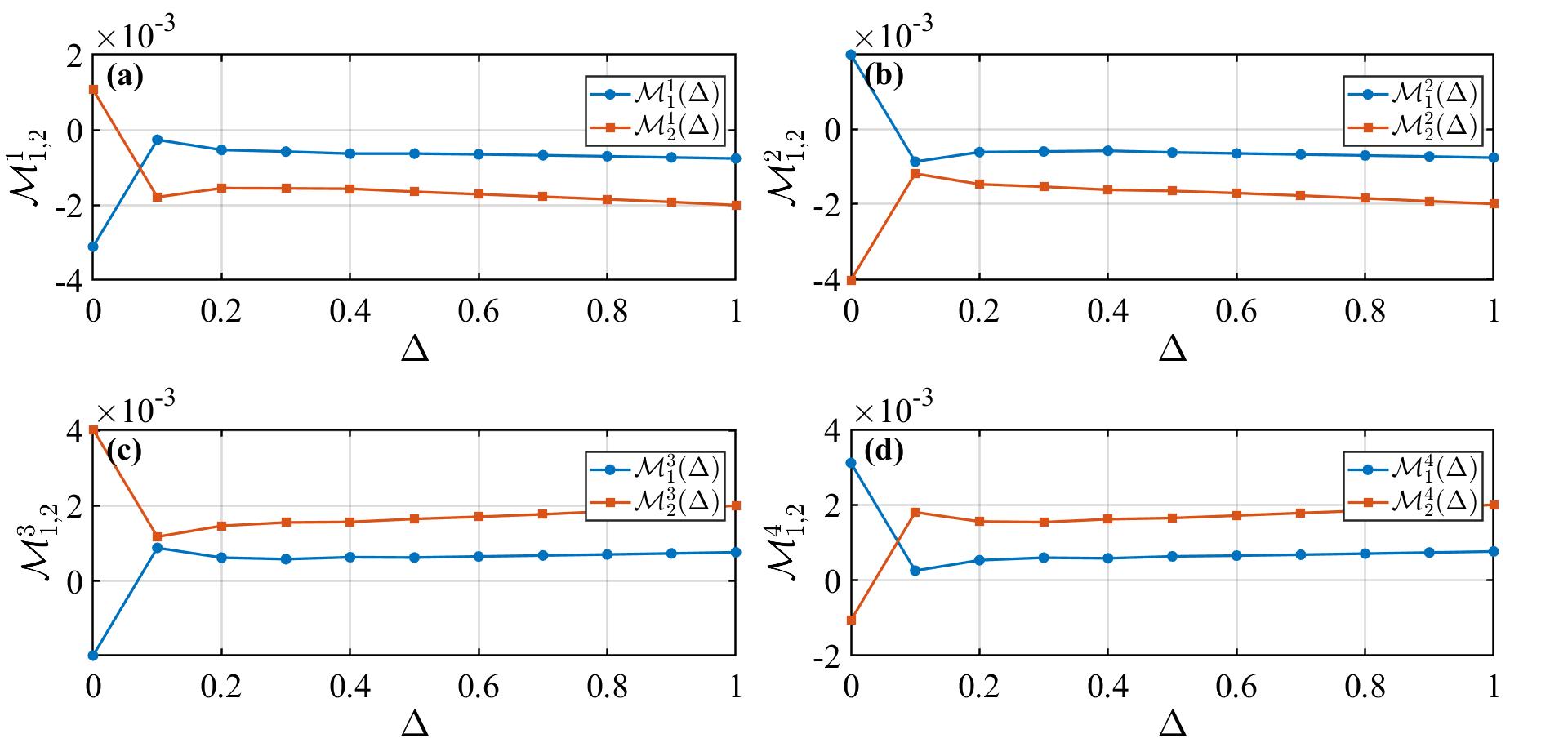}
    \caption{  {We show $\mathcal{M}^n_{1,2}=\frac{d\mathcal{E}_{1,2}^n(\Delta)}{dt}$ for $n=1,2,3$, and $4$ in (a,b,c), and  (d) for $m = 1$, respectively, as a function of strength of Wilson-Dirac mass term. One can observe  that $\mathcal{M}^1_{1,2}=-\mathcal{M}^4_{1,2}$ and $\mathcal{M}^2_{1,2}=-\mathcal{M}^3_{1,2}$. } 
    }
    \label{fig:slopes}
\end{figure}

\section{ {Role of Wilson-Dirac mass term in the energy exchange}}
\label{appC}

 {
To explicitly demonstrate the effect of the four-fold rotational symmetry-breaking term, we compute
the time-derivative of the instantaneous work done $\mathcal{M}^n_i=\frac{d\mathcal{E}_{i}^n(\Delta)}{dt}$ from $ H_{\mathrm{HOTI}}^{IC}(t)$ from $ H_{\mathrm{HOTI}}^{IC}(t)$, considering  $n=1,2,3,4$ as shown in Fig. \ref{fig:slopes} (a,b,c,d)  for both drive channels $i=1,2$, respectively.  }

 {
For $\Delta=0$, the slopes $\mathcal{M}^n_i$ corresponding to the two drives $i=1,2$ differ in sign irrespective of the band indices.
However, as soon as $\Delta>\Delta_t$,  $\mathcal{M}^{1,2}_i$ and $\mathcal{M}^{3,4}_i$ acquire negative and positive values, respectively for both the drive channels $i=1,2$. Note that $\Delta_t$ varies with $n$. Nonethelss, the magnitude of $\mathcal{M}^{n}_i$ changes with $\Delta>\Delta_t$, and there is a sharp change in their values at $\Delta \simeq 0.1$. Importantly, for $\Delta<\Delta_t$, $\mathcal{M}^{n}_1$ and  $\mathcal{M}^{n}_2$ are of opposite signs for all the bands. It is clear from the Fig. \ref{fig:slopes} that $\Delta (\cos(\omega_1 t + \phi_1)-\cos(\omega_2 t + \phi_2))T_4$ is responsible for emission or absorption of energy for both the drive channels given the same band index. Therefore, the identical sign of work done, associated with the two drive channels, is indeed caused by the TRS and  four-fold rotational symmetry breaking term when the amplitude of such term $\Delta$ is above a threshold value $\Delta_t$. This higher-order Wilson-Dirac mass  term gaps out the edge modes and hence counter-propagating channels cease to exist. The  identical sign of work done for both the drive channels can be attributed to a breakdown of counter-propagating modes. Interestingly,  the opposite signs of work done for the drive channels in FOTI phases can be considered as a manifestation of 
the counter-propagating modes along the boundaries.}

%

\end{document}